\documentclass[acmsmall,screen]{acmart}
\usepackage{makecell}
\usepackage{multirow}
\usepackage{algorithm}
\usepackage{algpseudocode}
\usepackage[most]{tcolorbox}

\renewcommand\footnotetextcopyrightpermission[1]{} 
\AtBeginDocument{%
  \providecommand\BibTeX{{%
    \normalfont B\kern-0.5em{\scshape i\kern-0.25em b}\kern-0.8em\TeX}}}

\begin{document}

\title{QuanTest: Entanglement-Guided Testing of Quantum Neural Network Systems}

\author{Jinjing Shi}
\orcid{0000-0002-0624-3340}
\affiliation{%
  \institution{School of Electronic Information, Central South University}
  \city{Changsha}
  \country{China}
  \postcode{410083}
}
\affiliation{%
  \institution{School of Computer Science and Engineering, Central South University}
  \city{Changsha}
  \country{China}
  \postcode{410083}
}
\email{shijinjing@csu.edu.cn}

\author{Zimeng Xiao}
\orcid{0009-0004-6842-5440}
\affiliation{%
  \institution{School of Computer Science and Engineering, Central South University}
  \city{Changsha}
  \country{China}}
  \postcode{410083}
\email{xiaozimeng@csu.edu.cn}

\author{Heyuan Shi}
\orcid{0000-0002-9040-7247}
\affiliation{%
  \institution{School of Electronic Information, Central South University}
  \city{Changsha}
  \country{China}}
  \postcode{410083}
\email{hey.shi@foxmail.com}
\authornote{Corresponding author.}

\author{Yu Jiang}
\orcid{0000-0003-0955-503X}
\affiliation{%
  \institution{School of Software, Tsinghua University}
  \city{Beijing}
  \country{China}}
  \postcode{100084}
\email{jy1989@mail.tsinghua.edu.cn}

\author{Xuelong Li}
\orcid{0000-0002-0019-4197}
\affiliation{%
  \institution{Institute of Artificial Intelligence (Tele AI), China Telecom}
  \city{Beijing}
  \country{China}}
  \postcode{100033}
\email{xuelong_li@ieee.org}
\renewcommand{\shortauthors}{Jinjing Shi, Zimeng Xiao and et al.}

\begin{abstract}

    Quantum Neural Network (QNN) combines the Deep Learning (DL) principle with the fundamental theory of quantum mechanics to achieve machine learning tasks with quantum acceleration. 
    Recently, QNN systems have been found to manifest robustness issues similar to classical DL systems. 
    There is an urgent need for ways to test their correctness and security. 
    However, QNN systems differ significantly from traditional quantum software and classical DL systems, 
    posing critical challenges for QNN testing. 
    These challenges include the inapplicability of traditional quantum software testing methods to QNN systems due to differences in programming paradigms and decision logic representations, 
    the dependence of quantum test sample generation on perturbation operators, 
    and the absence of effective information in quantum neurons. 
    In this paper, we propose QuanTest, a quantum entanglement-guided adversarial testing framework to uncover potential erroneous behaviors in QNN systems. 
    We design a quantum entanglement adequacy criterion to quantify the entanglement acquired by the input quantum states from the QNN system, 
    along with two similarity metrics to measure the proximity of generated quantum adversarial examples to the original inputs.
    Subsequently, QuanTest formulates the problem of generating test inputs that maximize the quantum entanglement adequacy and capture incorrect behaviors of the QNN system as a joint optimization problem and solves it in a gradient-based manner to generate quantum adversarial examples. 
    Experimental results demonstrate that QuanTest possesses the capability to capture erroneous behaviors in QNN systems (generating 67.48\%-96.05\% more high-quality test samples than the random noise under the same perturbation size constraints). 
    The entanglement-guided approach proves effective in adversarial testing,  generating more adversarial examples (maximum increase reached 21.32\%). 

\end{abstract}




\maketitle

\section{Introduction}

Quantum Neural Network (QNN) is a novel type of Deep Learning (DL) system that utilizes quantum computing as the underlying computational paradigm.
Driven by hybrid quantum-classical algorithms, QNN processes and computes quantum state data in quantum computers through Parameterized Quantum Circuits (PQCs) \cite{PQC}. 
A number of quantum learning algorithms leverage inherent properties such as quantum entanglement and quantum superposition \cite{QCQI}, 
exhibiting exponential advantages compared to their classical counterparts.
The Harrow-Hassidim-Lloyd (HHL) quantum algorithm \cite{HHL} and quantum principal component analysis \cite{PCA1, PCA2} are well-known examples that demonstrate this "quantum advantage."

Similar to classical Deep Neural Networks (DNNs), which still confront challenges related to robustness and security \cite{14ADV, FGSM}, 
QNN systems have also shown worrisome vulnerabilities in adversarial scenarios \cite{QAML1, QAML2}. 
This will pose significant challenges for the future applications of QNNs in industry, 
especially when deployed in safety and security-critical systems, 
where erroneous behaviors could lead to catastrophic events.
Therefore, the task of testing and evaluating QNN systems to uncover existing defects holds paramount importance in enhancing system security and quality.
Nonetheless, research on QNN system testing is currently missing.
The combination of quantum computing and deep learning brings a distinctive architecture to QNN systems, 
marked by substantial deviations from conventional quantum software and classical deep learning systems.
Consequently, there are critical challenges in the QNN systems testing and design of corresponding evaluation metrics.

{\bf The first challenge arises from the inapplicability of traditional quantum software testing methods to QNN systems.}
At present, quantum software testing has gradually attracted attention within the software engineering community \cite{QSE0}, 
and some quantum software testing methods targeting quantum programs have emerged recently \cite{QSE0.51, QSE0.52, QSE3, YMS}.
However, there is a fundamental distinction between traditional quantum software and QNN systems. 
At the physical level, a QNN system corresponds to a family of quantum circuits with variable structures and learnable parameters, rather than a fixed-scale specific circuit typically represented by traditional quantum software. 
At the logical level, a QNN system learns its rules autonomously from the features of data, whereas the system logic of traditional quantum software is directly specified by engineers. 
Therefore, these quantum software testing efforts for fixed circuits and decision logic of traditional software cannot be applied directly to QNN systems.

{\bf The second challenge is that the current testing methods and metrics for DL systems cannot be directly transferred to QNN systems.} 
While a number of testing methodologies \cite{Deepstellar, DLFuzz, Deeptest, Deepstate} have recently emerged for popular DL systems such as Convolutional Neural Networks (CNNs) and Recurrent Neural Networks (RNNs), 
along with a variety of testing metrics based on neuron coverage \cite{Deepxplore, Deepgauge}, surprise adequacy \cite{Guidingloss}, and loss \cite{Robot} have been proposed to appraise the quality of generated test cases, 
the quantum computing paradigm differs significantly from classical computing. 
Therefore, QNN testing is also different from classical DL testing as illustrated in Figure \ref{fig:1}. 
These differences are two-fold:
\begin{figure}[htbp]
  \centering
  \includegraphics[width=0.85\linewidth]{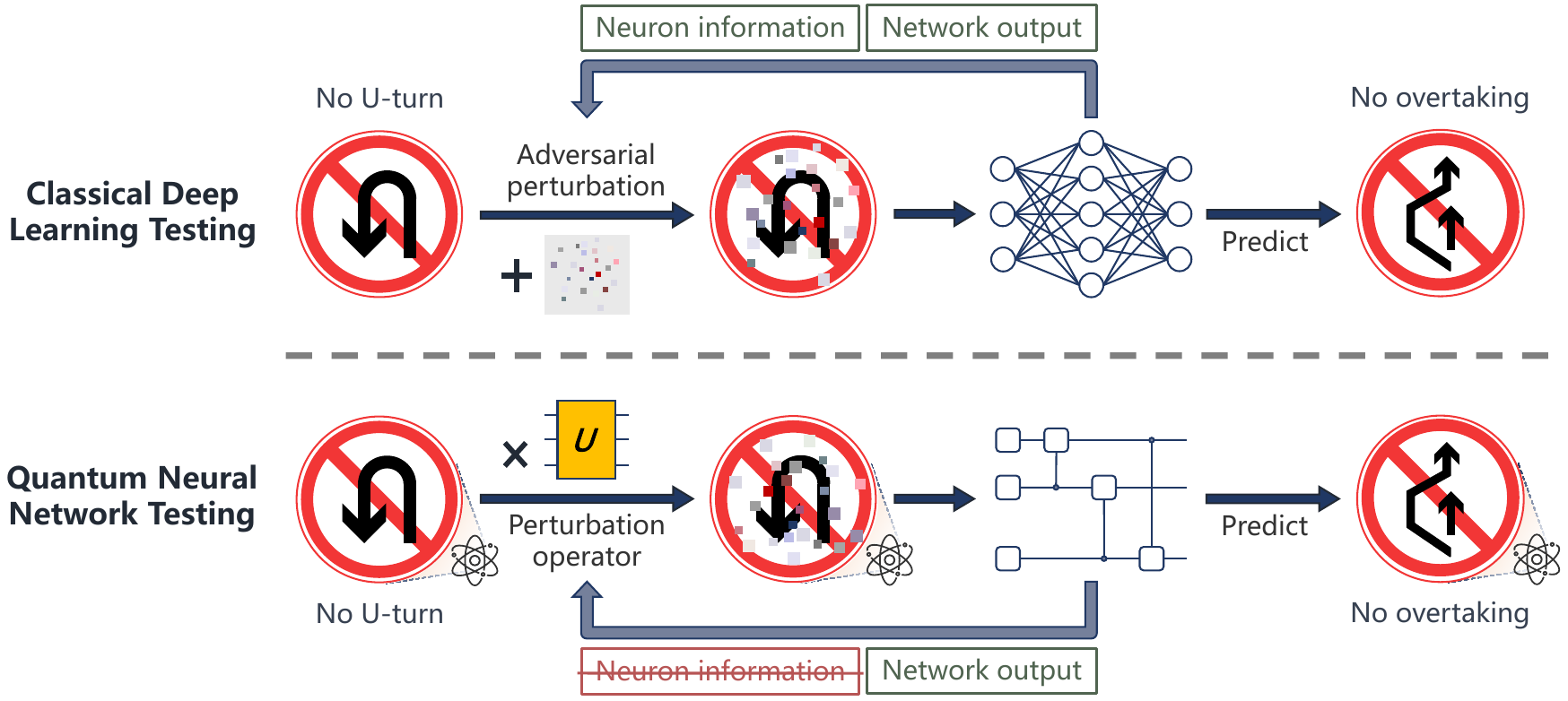}

  \caption{Differences between classical DL testing (upper part) and QNN testing (lower part). In QNN system testing, the QNN components cannot extract effective neuron information as in classical DNN, and the generation of adversarial examples is also different from the classical approach.}
  \Description{Differences between classical DL testing (upper part) and QNN testing (lower part). In QNN system testing, the QNN components cannot extract effective neuron information as in classical DNN, and the generation of adversarial examples is also different from the classical approach.}
  \label{fig:1}
\end{figure}

\begin{enumerate}
    \item {\bf The approach for obtaining quantum test samples diverges from that employed in the classical world.} 
    Due to the infinite-dimensional nature of Hilbert space, gathering manually labeled data to trigger every feasible logic in the QNN system is extremely impractical.
    Moreover, in the generation of adversarial examples, 
    classical computing can directly superimpose the original sample with the obtained perturbation in an additive manner to obtain the adversarial example. 
    In contrast, quantum computing typically involves applying a unitary operator layer, capable of causing perturbation effects, in the form of a function on the quantum state image.
    \item {\bf Limited useful information can be extracted from quantum neurons.}
    Owing to the high entanglement of quantum systems, apart from the parameters of rotation gates, 
    we cannot guarantee meaningful information about the QNN system from the output values of individual quantum neurons, nor can we mathematically provide a specific representation similar in meaning to classical neuron outputs \cite{QCQI}.
    Valuable information often concentrates on the overall output of QNN and the corresponding measurement results. 
    This renders existing DL testing metrics ineffective for QNN systems.
    Consequently, a novel evaluation metric tailored to quantum systems is imperative for QNN system testing.
  
\end{enumerate}

In this paper, we propose QuanTest, an adversarial testing framework for QNN systems guided by quantum entanglement, to address these challenges.
First, considering the unique quantum properties of the QNN system, 
we introduce the concept of quantum entanglement adequacy (QEA) to measure the test sufficiency of a set of test inputs to the QNN system based on the degree of activation of entanglement quantum neurons in QNN by the input quantum states.
Guided by quantum entanglement adequacy, QuanTest aims to generate adversarial examples that can trigger more solution spaces \cite{ExpEnt} in QNN, thereby uncovering more erroneous behaviors in the QNN system.
In particular, We first extract information from the quantum system constituted by the QNN model and quantum states. 
Considering the wave function collapse caused by observing the quantum systems \cite{QCQI}, 
in addition to measuring the output quantum states from the QNN model to obtain predictive information, 
it is necessary to retain some copies for subsequent testing to calculate the QEA and similarity metrics. 
For similarity evaluation, we propose two metrics, AFM and ATD, to measure the proximity between the generated adversarial examples and the original inputs.
Subsequently, we formulate the problem of generating test inputs that maximize quantum entanglement adequacy of a QNN system while also exposing as many erroneous behaviors as possible as a joint optimization problem.
We effectively solve this problem by obtaining perturbations in a gradient-based manner. 
The perturbations act directly on the original inputs in the form of a unitary operator, generating quantum adversarial examples through quantum evolution. 
Finally, we detect quantum adversarial examples and evaluate their quality through model performance and similarity metrics.

We implement QuanTest and conduct empirical studies on nine QNN systems, including three types each of QCL model \cite{mitarai2018quantum}, CCQC model \cite{schuld2020circuit}, and QCNN model \cite{cong2019quantum}.
These models were employed to perform three classification tasks based on two popular datasets MNIST \cite{lecun2010mnist} and Fashion-MNIST \cite{DBLP:journals/corr/abs-1708-07747}.
Specifically, we first evaluated the capability of QuanTest in quantum adversarial example generation.
The experimental results demonstrate that QuanTest can sensitively capture erroneous behaviors in QNN systems and generate high-quality quantum adversarial examples.
Compared with random coherent noise, QuanTest can generate adversarial examples that are 67.48\% to 96.05\% more than random coherent noise under the same similarity constraint with fidelity measure $>90\%$ or trace distance $<0.45$.
The performance of QuanTest is better in higher-dimensional quantum systems.
Furthermore, we assessed the effectiveness of entanglement guidance in adversarial testing through controlled experiments.
The results indicate that entanglement guidance can lead to the generation of adversarial test inputs that trigger more solution space in QNN systems.
This capability is related to the proportion of the impact of entanglement in the output quantum state in QEA.

The main contributions of this paper are summarized as follows:
\begin{itemize}
	\item We introduce quantum entanglement adequacy as the first test adequacy metric for QNN systems, and experimentally demonstrated the effectiveness of entanglement guidance in quantum adversarial testing.

	\item We propose and implement the first quantum adversarial testing framework, QuanTest\footnote{https://github.com/am0x00/QuanTest}. QuanTest formulates the problem of both maximizing the quantum entanglement adequacy and the number of incorrect behaviors as a joint optimization problem. This is effectively solved by iteratively updating the perturbation operator.
	\item We evaluate the effectiveness of QuanTest through numerical simulations on nine QNN systems. The experiment results demonstrate that QuanTest can effectively detect erroneous behaviors in QNN systems and generate high-quality quantum adversarial examples under the guidance of quantum entanglement adequacy.
\end{itemize}

\section{Background}

\subsection {Preliminary of Quantum Computing}
\subsubsection {Qubits}
Qubits are the quantum counterpart to classical bits in the realm of quantum computing, serving as the fundamental units of quantum computation. 
Unlike classical bits, which can only exist in states of 0 and 1, qubits possess unique characteristics. 
They can not only be in states denoted as $|0\rangle$ and $|1\rangle$ ("$| \rangle$" represents the Dirac notation), but they can also exist in superpositions of these two states. 
For example, a quantum state $|\psi\rangle$ on a Hilbert space $\mathbb{C}^{2}$ can be represented as:
\begin{equation}
	|\psi\rangle=\alpha|0\rangle+\beta |1\rangle,
\end{equation}
where $\alpha$ and $\beta$ are complex numbers, representing probability amplitudes and satisfying the condition $|\alpha|^2 + |\beta|^2 = 1$. 
$\{|0\rangle,|1\rangle\}$ consists of two orthogonal 2-dimensional state vectors and is known as the computational basis in this Hilbert space.
A composite system of $n$ qubits can be described by a unit complex vector in a $2^n$-dimensional Hilbert space. 
Note that the quantum state space of the composite system is the tensor product of its component system's state spaces.
\subsubsection {Quantum Logic Gates}
The construction of quantum gates is fundamentally different from classical logic gates. 
In quantum circuits, any quantum gate can correspond to a unitary operator $U$, which is mathematically represented by a unitary matrix in terms of logic.
The unitary matrix $U$ satisfies the property of unitarity: 
\begin{equation}
	U^{\dagger} U=UU^{\dagger}=I,
\end{equation}
where $U^{\dagger}$ is the conjugate transpose of $U$, and $I$ represents the identity matrix. 
This implies that all quantum gates must be reversible, and the dimensions of input and output must remain consistent to preserve the unitary transformation.
A quantum gate acting on $n$ qubits $|\psi\rangle$ can be described by a $2^n \times 2^n$ unitary matrix $U$.
The essence of this operation is to perform matrix operations on the state vector, i.e., $U|\psi\rangle$.
Common quantum logic gates can be categorized based on their properties and functions into fixed gates, rotation gates, and controlled gates, among others.
\subsubsection {Quantum Measurement}
To obtain information about the internal state of a quantum system, quantum measurements are often necessary. 
Measurements in quantum computing do not exist as quantum logic gates since they are irreversible. 
When we perform measurements on a closed quantum system, the system is no longer in a closed state, meaning it no longer follows unitary evolution. 
At this point, the quantum state randomly collapses onto the basis states corresponding to the Hermitian operator M associated with the measurement. 
Therefore, to obtain meaningful measurement results, measurements often need to be repeated multiple times.
\subsubsection {Quantum Entanglement}
In a composite system, a quantum state that cannot be written as the tensor product of its component systems states is referred to as an entangled state.
Entanglement can be generated through the interaction between particles, a property that is independent of the distance between the particles. 
In quantum computing, entanglement is often created by applying controlled gates (including CNOT, iSWAP, etc.) to qubits. Bell states (such as $(|{00}\rangle+|{11}\rangle)/\sqrt{2}$) are the maximally entangled quantum states of two qubits \cite{QCQI}. 
Entanglement is a unique resource of quantum mechanics that plays a crucial role in quantum computing and quantum information.

\begin{figure}[tbp]
  \centering
  \includegraphics[width=0.6\linewidth]{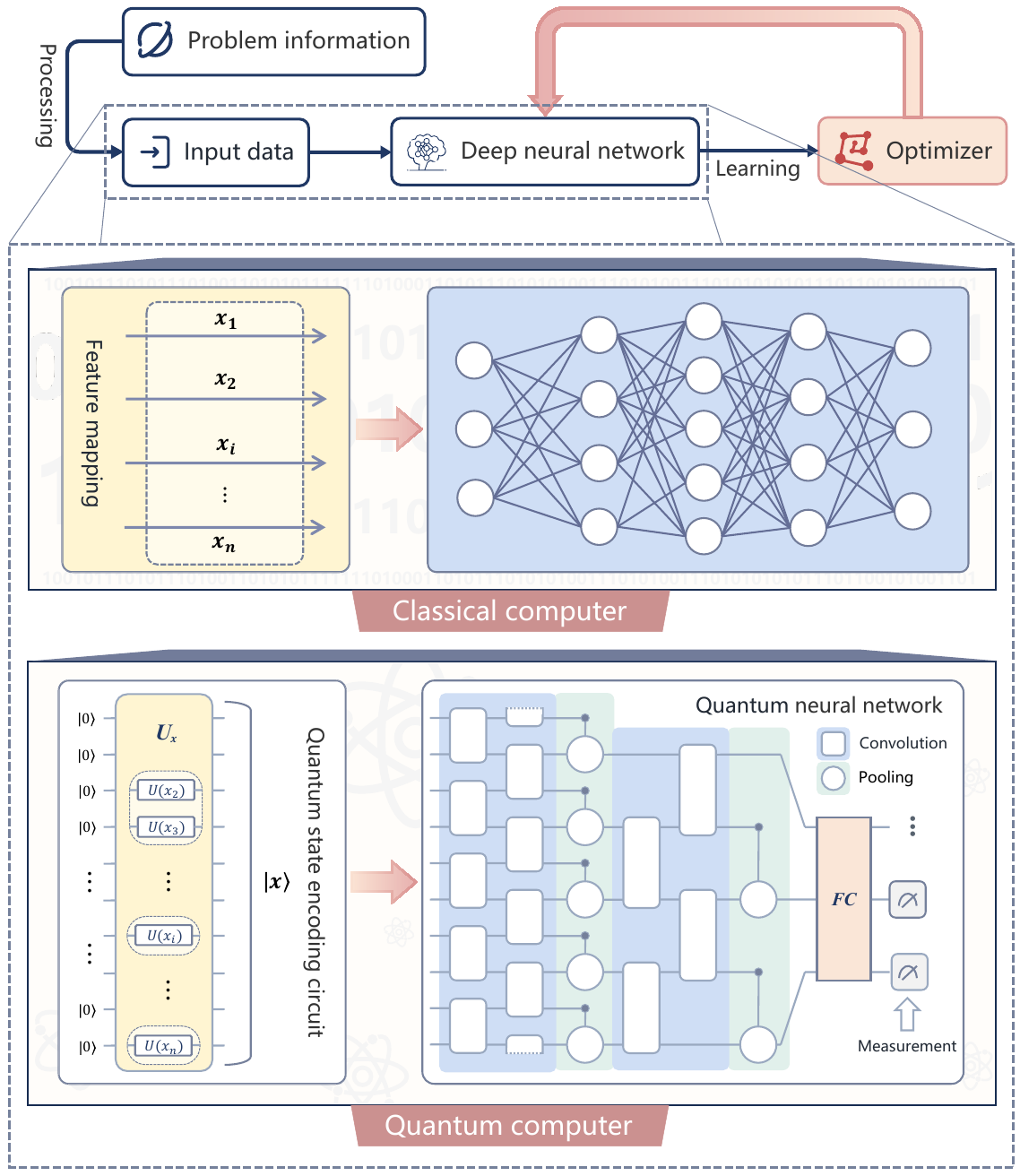}
  \caption{Differences between QNN and classical DNN during the training process.}
  \Description{Differences between QNN and classical DNN during the training process.}
  \label{fig:2}
\end{figure}

\subsection{Quantum Neural Network}
Quantum neural networks (QNNs) are a type of neural network model based on quantum computing, 
with the core being parameterized quantum circuits (PQCs) \cite{PQC} driven by variational quantum algorithms \cite{VQA}. 
Similar to classical feedforward neural networks, 
QNNs learn by updating the trainable parameters of rotation gates within the circuits through gradients——this aspect is executed by classical optimizers, 
and the processing and computation of the target quantum state are carried out by quantum computers. 
An example \cite{QCNNimg} of using a QNN model $\mathcal{M}(|\boldsymbol{x}\rangle; U_{\boldsymbol{\Theta}})$ for machine learning tasks is shown in Figure \ref{fig:2}. 
The training data $\boldsymbol{x}$ is first encoded as quantum states $|\boldsymbol{x}\rangle$.
Subsequently, these states are input into the QNN for computation. 
The trainable parameters $\boldsymbol \Theta$ within the PQC $U_{\boldsymbol{\Theta}}$ are updated using classical optimizers until the results meet the expected termination criteria.
In this process, the PQC $U_{\boldsymbol{\Theta}}$ can be regarded as a unitary operator, and its action on the quantum state $|\boldsymbol{x}\rangle$ can be represented as $U_{\boldsymbol{\Theta}}|\boldsymbol{x}\rangle$. The final output of the QNN can only be obtained after measuring $U_{\boldsymbol{\Theta}}|\boldsymbol{x}\rangle$.
Note that the data learned by QNN can be either classical data or quantum data, with classical data ends up encoded as quantum states \cite{QML}. 
The process of quantum state encoding can be viewed as a mapping from classical dimensions to quantum dimensions, so the direct input of QNNs is always quantum states

As described in Section 2.1, the operations performed by quantum neurons as quantum gates are linear and unitary. 
Therefore, although the learning framework for performing tasks is similar, there are still many differences between the construction of QNNs and feedforward neural networks: 
the number of quantum neurons in each layer of the network is strictly limited by the number of qubits, 
and due to the noise in noisy intermediate-scale quantum (NISQ) devices, the scale of QNNs that current quantum computing devices can load is often limited in depth and width.
Currently, QNN has given rise to various variants, including quantum convolutional neural networks (QCNNs) \cite{cong2019quantum} and quantum recurrent neural networks (QRNNs) \cite{QRNN}, among others. 
Novel network architectures are still being continuously proposed.

\section{Design}

\begin{figure*}[tb]
	\centering

    \includegraphics[width=\columnwidth]{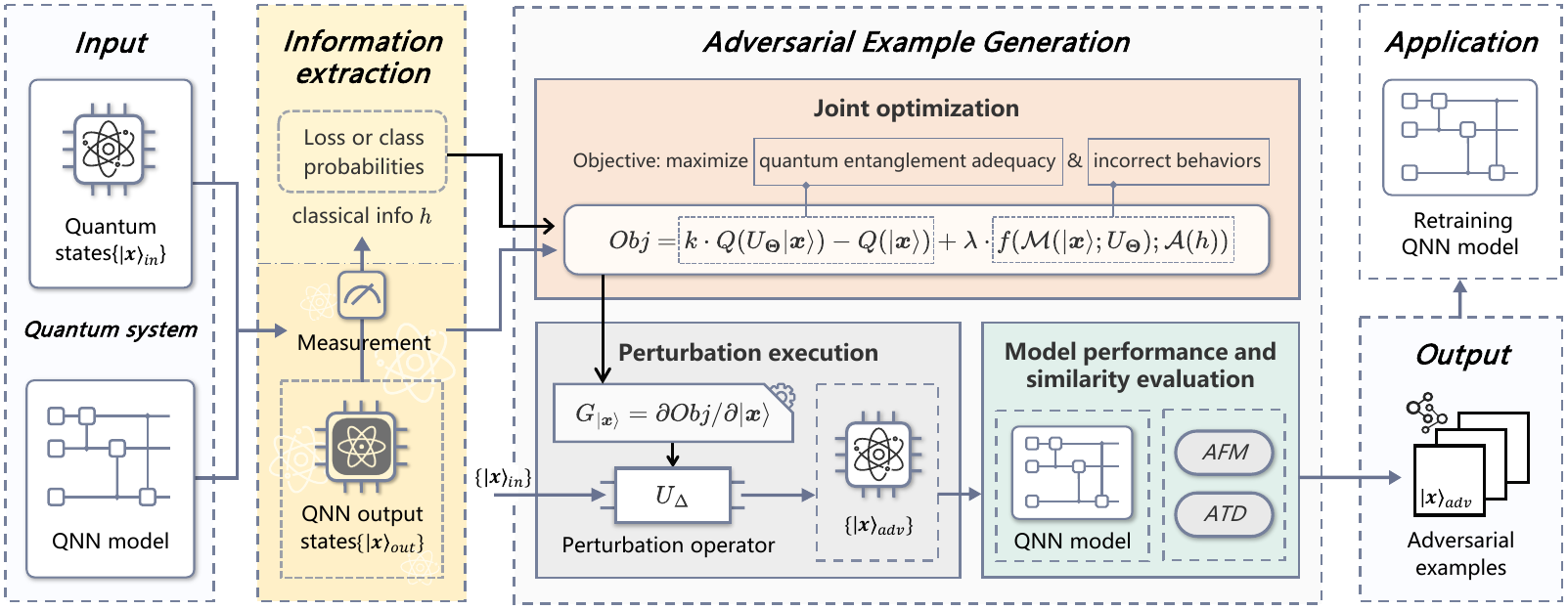}

	\caption{Overview of the QuanTest framework. QuanTest takes a quantum system composed of quantum states containing classical input information and a pre-trained QNN model as input. It extracts information generated by the quantum system.
    Based on this information, QuanTest is capable of generating quantum test samples that are adequately entangled with the QNN while inducing erroneous behaviors in the QNN system. In the diagram, black open arrows refer to the flow of classical information while the gray-blue arrows refer to the flow of quantum information.}
    \Description{Overview of the Quantest framework. QuanTest takes a quantum system composed of quantum states containing classical input information and a pre-trained QNN model as input. It extracts information generated by the quantum system.
    Based on this information, QuanTest is capable of generating quantum test samples that are adequately entangled with the QNN while inducing erroneous behaviors in the QNN system. In the diagram, black open arrows refer to the flow of classical information while the gray-blue arrows refer to the flow of quantum information.}
	\label{fig:3}
\end{figure*}

In this section, we provide an overview of our adversarial testing framework, QuanTest. Figure \ref{fig:3} shows the workflow of QuanTest.

In the information extraction module, QuanTest utilizes quantum measurements to extract information generated during the quantum evolution of quantum states in the QNN model. 
For instance, classical information such as loss and class probabilities, which are used for adversarial attacks, needs to be measured from the quantum states outputted by the QNN model and then obtained through computation. 
In addition, in actual experiments, the calculation of a series of metrics designed based on quantum properties, such as the entanglement degree of quantum states and measures of similarity between quantum states, also often requires measurements of the quantum states.
Considering the inherent closed nature of quantum systems that inevitably leads to the collapse of the wave function upon direct observation of quantum states \cite{QCQI}, at this stage, QuanTest additionally retains a portion of the copies of quantum states input and output from the QNN model for subsequent calculations of QEA and similarity evaluation. 

Adversarial example generation is the core component of QuanTest, comprising three modules: joint optimization, perturbation execution, and quality evaluation for adversarial examples. 
QuanTest increases the model loss or the probability of target classes based on the performed quantum-adapted adversarial attack strategy, enabling QuanTest to search for adversarial examples in a lightweight and scalable manner.
Meanwhile, QuanTest also aims to enhance the entanglement of generated adversarial examples during the testing process for a more comprehensive coverage of the solution space of QNN system. 
Specifically, QuanTest combines the maximization of incorrect behaviors and QEA into a joint optimization problem, employing a gradient-based algorithm for effective solution. 
Note that, unlike typical metrics such as neuron coverage, the calculation of QEA does not originate from classical information with gradients directly extractable from QNN models. 
Moreover, the measure of quantum entanglement is complex and non-unique, making the design of a suitable metric for assessing entanglement adequacy crucial.
We discuss key insights into the entanglement adequacy in the subsequent Section 4.2.

Based on the perturbations obtained from the solution, QuanTest independently modifies each component of the input state vector and normalizes the modified input to ensure that it still represents a quantum state in the quantum world. 
In the quantum circuit, this modification takes the form of a perturbation operator directly acting on the original input quantum state. 
Without the need for additional domain constraints, the original input can evolve into quantum adversarial examples through the perturbation operator. 
The construction of the perturbation operator is not rigid; it can be any unitary transformation close to the identity operator. 
Finally, these quantum adversarial examples are evaluated based on the performance of the tested model and similarity to the original samples to ensure the quality of the testing. 
The generated adversarial example set can be used for applications such as retraining to enhance the robustness of the tested model. 

\subsection{Quantum Entanglement Adequacy Metrics}

In classical DL systems, acquiring the weights and outputs of each neuron within the DNN is straightforward. 
However, QNNs differ in this regard. 
Quantum neurons are essentially quantum logic gates represented in the form of unitary operators, 
making it impossible to access the output values of any individual quantum neuron in the network. 
This implies it is currently challenging to select specific quantum neurons and judge their coverage in the testing. 
To intuitively measure the adequacy of QNN system testing and better guide the generation of test samples, 
we consider introducing the concept of entanglement in quantum computing. 

One core reason quantum computation can demonstrate potential quantum advantage in many problems is that it can construct an exponentially large quantum state space through controllable entanglement and interference \cite{havlivcek2019supervised}.
By utilizing low-depth circuits to generate quantum states with high entanglement characteristics, QNN models may produce atypical patterns that classical models cannot effectively produce.
These patterns can efficiently represent solution space for tasks such as prediction and classification, capturing nontrivial correlations in the data \cite{schuld2020circuit}.
Quantum machine learning models such as quantum circuit learning (QCL) \cite{mitarai2018quantum}, parameterized Hamiltonian learning (PHL) \cite{PHL}, and quantum kernel methods \cite{QKM} achieve strong entanglement in quantum circuits for enhancing the entanglement capability of QNNs by incorporating various configurations of multi-qubit gates such as CNOT, iSWAP, and their extended variants. 
We can perceive that, as the quantum state used as a test input undergoes computation within a QNN, 
the more fully it activates the entangled quantum neurons, 
the higher the entanglement degree it obtained from the QNN.
This leads to a more comprehensive coverage of the solution space in the QNN.
Therefore, we propose a method to calculate the QEA of QNNs.
This is achieved by measuring the entanglement of both the test input quantum state and the resulting output quantum state after undergoing unitary operations within the QNN. 
The difference in measurement outcomes is used to represent the QEA of that particular test case.

\begin{definition} [Quantum entanglement adequacy, QEA] 
Given a fully trained QNN model $\mathcal{M}$ and a test input set $\mathcal{T}=\{\boldsymbol{x_1},\boldsymbol{x_2},\ldots\}$, 
the action of $\mathcal{M}$ on the input quantum state $|\boldsymbol{x}\rangle$ can be equivalently represented as the multiplication of $|\boldsymbol{x}\rangle$ and a unitary matrix $U_{\boldsymbol{\Theta}}$,
where $|\boldsymbol{x}\rangle$ represents the quantum state that encodes the information of $\boldsymbol{x} \in \mathcal{T}$,
and $\boldsymbol{\Theta}$ represents the learned and fixed gate parameters in $\mathcal{M}$. 
Note that the result output of $U_{\boldsymbol{\Theta}}|\boldsymbol{x}\rangle$ is a quantum state with the same dimension as $|\boldsymbol{x}\rangle$.
Then, the quantum entanglement adequacy (QEA) can be expressed as:
\begin{equation}
	QEA(\mathcal{T}, \mathcal{M}, k) = 
	\frac1{|\mathcal{T}|} \sum_{x_i \in \mathcal{T}} 
	\big\vert k \cdot Q(U_{\boldsymbol{\Theta}}|\boldsymbol{x_i}\rangle)-Q(|\boldsymbol{x_i}\rangle)\big\vert,
\end{equation}
where $Q(|\boldsymbol{x}\rangle)$ represents the MW entanglement measure for $|\boldsymbol{x}\rangle$, and $k$ is used to adjust the weight of $Q(U_{\boldsymbol{\Theta}}|\boldsymbol{x}\rangle)$, 
in order to avoid imbalance in QEA calculations caused by the concentration of $Q(U_{\boldsymbol{\Theta}}|\boldsymbol{x}\rangle)$ at higher values when the entanglement capability \cite{ExpEnt} of QNN is strong. 
\end{definition}

\begin{figure*}[tp]
	\centering
    \includegraphics[width=0.77\linewidth]{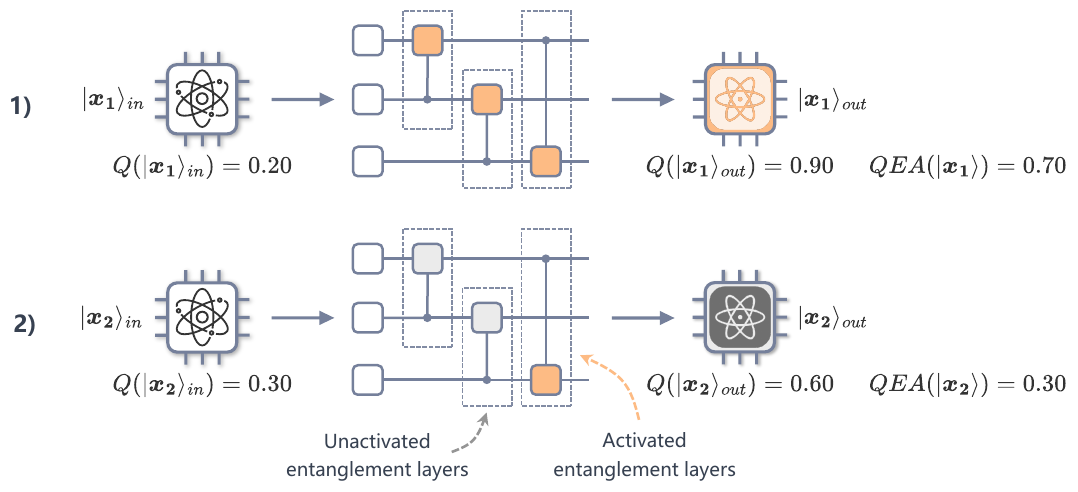}
	\caption{Example of QEA computation for two quantum states in the QNN. The nodes in gray denote quantum entanglement layers that are not sufficiently activated, while aqua nodes represent those that are adequately activated.}
    \Description{Example of QEA computation for two quantum states in the QNN. The nodes in gray denote quantum entanglement layers that are not sufficiently activated, while aqua nodes represent those that are adequately activated.}
	\label{fig:3.5}
\end{figure*}

For the specific calculation of $Q$, there are several methods currently available for quantifying entanglement as a resource. 
Considering that the selected measurement approach should possess a certain degree of scalability and computational simplicity, to facilitate the formulation of a guiding optimization objective,  
in this paper, we adopt a polynomial global measure for multipartite entanglement, known as the Meyer-Wallach (MW) entanglement measure \cite{entanglement}. 
The MW entanglement measure is defined as follows. For an $n$-qubit quantum system on the Hilbert space $\mathbb{C}^{2^n}$, define a linear map $\iota_j(b)$ that acts on the computational basis $|b_1\ldots b_n\rangle$: 
\begin{equation}
	\iota_j(b)|b_1\ldots b_n\rangle = \delta_{bb_j}|b_1\ldots \widehat{b_j} \ldots b_n\rangle,
\end{equation}
where $b \in \{0,1\}$, $\widehat{b_j}$ denotes the absence of the $j$-th qubit,
and $\delta_{ij}$ is the Kronecker delta. 
Through the map, a new Hilbert space $\mathbb{C}^{2^{n-1}}$ can be obtained. 
For $|{u}\rangle,|{v}\rangle \in \mathbb{C}^{2^{n-1}}$ with $|{u}\rangle=\sum u_i|{i}\rangle$ and  $|{v}\rangle=\sum v_i|{i}\rangle$, the area of the parallelogram $S$ they form in this Hilbert space is defined as:
\begin{equation}
	S(|u\rangle,|v\rangle)=\sum_{i<j}|u_iv_j-u_jv_i|^2.
\end{equation}
Finally, for $|{x}\rangle \in \mathbb{C}^{2^n}$, the MW entanglement measure Q is:
\begin{equation}
	Q(|{x}\rangle)=\frac4n\sum_{j=1}^nS\big( \iota _j\left(0\right)|{x}\rangle, \iota _j\left(1\right)|{x}\rangle\big),
\end{equation}
where $4/n$ factor serves the purpose of normalization. 
The value range of $Q(|{x}\rangle)$ is $[0,1]$. 
The larger the value of Q, the higher the degree of entanglement possessed by the quantum state.
For instance, in the case of a pure state $|{\phi_1}\rangle = |{00}\rangle$, $Q(|{\phi_1}\rangle) = 0$, indicating the absence of entanglement. On the other hand, for a Bell state $|{\phi_2}\rangle = (|{00}\rangle+|{11}\rangle)/\sqrt{2}$,  $Q(|{\phi_2}\rangle) = 1$, signifying complete entanglement.

We show the computation process of QEA using Figure \ref{fig:3.5} as an example. 
Quantum state data undergo entanglement measurements during the input and output of the QNN. 
The more entangled neurons activated by the quantum state, the greater the entanglement properties obtained. 
The difference between entanglement measurements before and after, i.e., the QEA, is consequently higher.

\subsection{QuanTest Algorithm}

In this section, we describe the details of how QuanTest solves the joint optimization problem of maximizing incorrect behaviors and QEA through gradient ascent to generate adversarial examples as test inputs.
Guided by entanglement, the adversarial examples generated by QuanTest are expected to trigger a broader solution space within the QNN system to discover more erroneous behaviors.

The specialized characteristics of quantum computation necessitate that QuanTest constantly considers a good fit with quantum systems in algorithm implementation and metric design, ensuring efficient execution of scalable automated testing on quantum computing devices. 
Algorithm \ref{algorithm1} presents the core process of adversarial example generation guided by entanglement. 
Below, we discuss the specific details of the algorithm and solutions to relevant quantum challenges around three key components of the algorithm.

\subsubsection{Joint Optimization} For most classical deep learning systems, the gradient of the objective function can be efficiently computed when the weight parameters and intermediate neuron values of the pre-trained DNN are transparent. 
QNN systems based on the same classical optimizer also share similar characteristics in gradient computation.
As described in Section 2.2, QNN completes machine learning tasks by updating the learnable parameters of quantum logical gates in the PQC. 
Since QuanTest has access to the parameters of quantum neurons in the QNN, it can efficiently compute the gradient of the optimization objective when the parameters of the PQC are constants and the input quantum state is variable.

The optimization objective of QuanTest is defined by the following equation:
\begin{equation}
	Obj=k \cdot Q(U_{\boldsymbol{\Theta}}|\boldsymbol{x}\rangle)-Q(|\boldsymbol{x}\rangle)+\lambda \cdot f(\mathcal{M}(|\boldsymbol{x}\rangle;U_{\boldsymbol{\Theta}});\mathcal{A}).
\end{equation}
It consists of two parts, where the first part $k \cdot Q(U_{\boldsymbol{\Theta}}|\boldsymbol{x}\rangle)-Q(|\boldsymbol{x}\rangle)$ is to generate test inputs that maximize QEA, and the second part $f(\mathcal{M}(|\boldsymbol{x}\rangle; U_{\boldsymbol{\Theta}});\mathcal{A})$ is to generate test inputs that trigger as many incorrect behaviors in the QNN as possible. The hyperparameter $\lambda$ is used to balance the two optimization objectives during the joint optimization process. 
\begin{center}  
\begin{minipage}{\linewidth}  
\begin{algorithm}[H]
  \caption{Adversarial Example Generation Guided by Entanglement}
  \label{algorithm1}
    \begin{algorithmic}[1]
    \Require
			inputs $\gets$ classic original inputs
			
			qnn $\gets$ QNN model under test
			
			strategy $\gets$ strategy for quantum adversarial example generation

			w $\gets$ weight of entanglement adequacy in the joint goal    
			
			k $\gets$ parameter to balance the entanglement of input states and output states
			
			r $\gets$ step size in gradient ascent

    \Ensure
    adversarial test input set $\mathbf{S_{adv}}$

    \State $\mathbf{S_{adv}}$ = []
    \For{each $\boldsymbol{x}\in $ inputs}
        \State $|\boldsymbol{x}\rangle$ = encoding($\boldsymbol{x}$); {\small{\it {//quantum state encoding}}}
        \State $|\boldsymbol{x}\rangle_{adv}$ = encoding($\boldsymbol{x}$); {\small{\it {//copy of $|\boldsymbol{x}\rangle$ for subsequent quantum adversarial example generation}}}
		\State $y_{ori}$ = qnn($|\boldsymbol{x}\rangle$).measurement(); {\small{\it {//prediction lable obtained by observing quantum system}}}
        \While{True}
            \State $|\boldsymbol{x}\rangle_{out}$ = qnn($|\boldsymbol{x}\rangle_{adv}$); {\small{\it {//  output quantum state}}}
            \State obj = []; 
			\State obj.append(obj\_incorrect($|\boldsymbol{x}\rangle_{out}$, strategy));
			\State obj.append(w $\cdot$ obj\_entanglement($|\boldsymbol{x}\rangle_{adv}$, $|\boldsymbol{x}\rangle_{out}$, k));
			\State $G_{|\boldsymbol{x}\rangle_{adv}}$ = r $\cdot$ $\partial \text{obj} / \partial |\boldsymbol{x}\rangle_{adv}$; {\small{\it {//perturbation operator with gradient information}}}
			\State $|\boldsymbol{x}\rangle_{adv}$ = PERTURBATION\_OP($|\boldsymbol{x}\rangle_{adv},  G_{|\boldsymbol{x}\rangle_{adv}}$); {\small{\it {//quantum adversarial examples obtained}}}
			\State $s$ = measure\_similarity($|\boldsymbol{x}\rangle$, $|\boldsymbol{x}\rangle_{adv}$);
			\State $y_{adv}$ = qnn($|\boldsymbol{x}\rangle_{adv}$).measurement();
			\If {$y_{adv}$ != $y_{ori}$ and $s$ meets the expectation} 

				\State $\mathbf{S_{adv}}$.append($|\boldsymbol{x}\rangle_{adv}$);
				\State \textbf{break};

			\EndIf  
        \EndWhile
	\EndFor
	\Function{perturbation\_op}{$|\boldsymbol{x}\rangle_{adv},  G_{|\boldsymbol{x}\rangle_{adv}}$}
	    \For{each ${x}_i\in |\boldsymbol{x}\rangle_{adv}$}
            \State $\Delta = {x_i} + G_{{x_i}}$
            \State ${x}_i^* = {x}_i + \Delta$
        \EndFor
        \State \Return $|\boldsymbol{x^*}\rangle_{adv}$
    \EndFunction
  \end{algorithmic}
\end{algorithm}
\end{minipage}
\end{center}
We treat these two parts as a joint optimization problem and maximize this objective function (Algorithm \ref{algorithm1} line 8-11).

\begin{itemize}
\item {\bf Maximizing quantum entanglement adequacy.}
QuanTest utilizes the proposed QEA metric to guide the generation of adversarial examples, thereby more comprehensively exploring the solution space of QNN. 
As described in Section 3.2, due to the difficulty in determining the specific state of a single quantum neuron, we assess the adequacy of the entanglement obtained from the QNN by separately measuring the entanglement degree of the quantum state before and after passing through the QNN.
Unlike optimization objectives such as neuron coverage, which directly reflect the model's state and are easy to obtain gradients, QEA indirectly captures the state of the QNN model based on the nature changes of quantum data, and the calculation of entanglement metric $Q$ is more complex. 
The favorable properties of the MW entanglement measure ensure that QuanTest can effectively compute gradients for QEA, allowing QuanTest to directly use QEA as an optimization objective.

\item {\bf Maximizing incorrect behaviors.}
Modular construction enables QuanTest to conveniently select different quantum-adapted adversarial attack strategies $\mathcal{A}$. 
The determination of this strategy dictates the prediction information to be further calculated after measuring the output quantum state during the information extraction. 
For instance, the quantum-adapted FGSM strategy requires providing loss, while the quantum-adapted DLFuzz strategy needs category probabilities as input.
In this part, QuanTest aims to modify the input in the form of adversarial attacks, leading the model to exhibit incorrect behavior.
\end{itemize}

All terms of the jointly optimized objective are differentiable. 
QuanTest can iteratively modify the input $|\boldsymbol{x}\rangle$ by using gradient ascent to maximize these terms. 
The computed gradient $\frac{\partial \text{obj}}{\partial|\boldsymbol{x}\rangle}$ will become the perturbation, which is subsequently transformed into a perturbation operator on a quantum circuit for generating adversarial examples.

\subsubsection{Perturbation Execution}

Due to its fundamentally different computational paradigm compared to classical computing, the original quantum state cannot be simply superimposed with perturbations in an additive manner to obtain adversarial examples. 
This direct superposition has the potential to disrupt the unitarity of the quantum system.
In practice, QuanTest independently modifies each component in the state vector and normalizes the modified state vector when necessary to ensure that it still represents a quantum state in the quantum world (Algorithm \ref{algorithm1} line 12 and function $\tt PERTURBATION\_OP$). 
In quantum circuits, this perturbation can be achieved by constructing a perturbation operator that acts directly on the original quantum state and has a sufficiently large search space, as shown in the lower half of Figure \ref{fig:1}. 
The perturbation operator does not have a fixed construction form; a simple local unitary transformation-based perturbation operator may not have a sufficiently large search space to find the target solution, while complex global operations require extremely high precision and cost. 
Additionally, the perturbation operator is constrained to approach the identity operator to maintain the perturbation reasonably small.

\subsubsection{Quality Evaluation} 

The generated quantum adversarial examples need to undergo quality evaluation in terms of model performance and similarity before being output as test samples (Algorithm \ref{algorithm1} line 13-15). 
A qualified test sample should first be able to induce erroneous behavior in the model, 
such as causing a change in the predicted labels of the classification model.
Regarding the similarity evaluation of image data that has already been encoded into quantum states, 
we cannot directly employ classical methods to measure the difference or closeness between two input vectors. 
Serving as quantum generalizations of the classical notions, fidelity \cite{Fidelity} and trace distance \cite{Trace} are two frequently utilized distance measures for quantum information. 
They have also been used in some quantum machine learning work \cite{QAML1, QAML12, QAML13} to evaluate the similarity between quantum states. 
Here, we choose to use these measures to calculate perturbations that fool the QNN system.
As a result, we introduce the definitions of Average Fidelity Measure (AFM) and Average Trace Distance (ATD) to quantitatively assess the similarity between the original input quantum state and the quantum test samples, 
thereby evaluating the quality of the generated test samples.
In fact, these metrics can also be effectively employed to assess the robustness of the model.

\begin{figure}[htbp]
  \centering
  \includegraphics[width=0.7\linewidth]{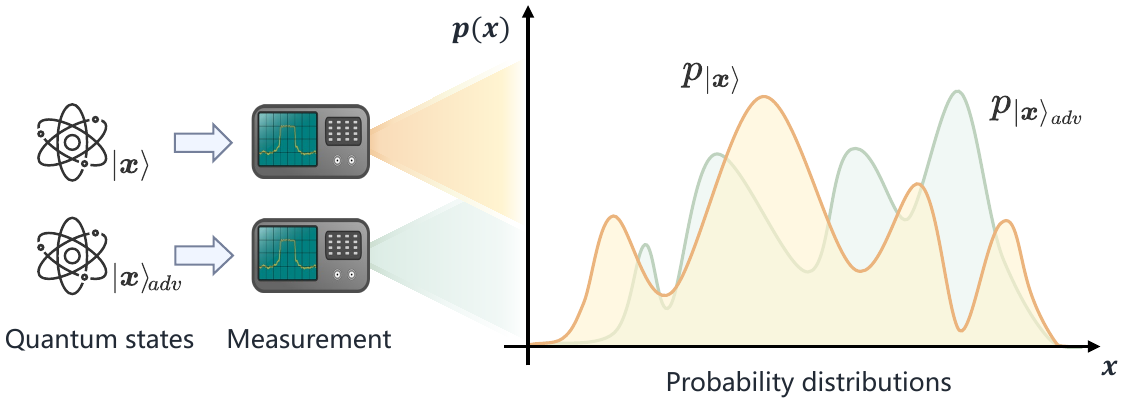}
  \caption{Illustration of fidelity visualization between the original quantum sample $|\boldsymbol{x}\rangle$ and quantum adversarial example $|\boldsymbol{x}\rangle_{adv}$. The light orange and light green curves represent the probability distributions $p_{|\boldsymbol{x}\rangle}$ and $p_{|\boldsymbol{x}\rangle_{adv}}$ obtained through POVM measurements for the original sample and adversarial example, respectively.}
  \Description{Visualization of fidelity between two quantum states based on POVM measurements.}
  \label{fig:4}
\end{figure}

\begin{definition} [Average fidelity measure, AFM] 
We obtain the AFM by calculating the sample mean of fidelity between the original inputs in $\mathcal{T}$ and their generated test samples. 
More formally: 
\begin{equation}
	AFM(\mathcal{T}) =\frac1{|\mathcal{T}|} \sum_{x_i \in \mathcal{T}} F(|\boldsymbol{x}_i\rangle\langle\boldsymbol{x}_i|,|\boldsymbol{x}_i\rangle_{adv}\langle\boldsymbol{x}_i|_{adv}),
\end{equation}
where $|\boldsymbol{x}_i\rangle_{adv} = U_{\psi}|\boldsymbol{x}_i\rangle$, 
$U_{\psi}$ is a perturbation operator acting on  $|\boldsymbol{x}_i\rangle$, 
and $F(\rho, \sigma)$ is the fidelity between quantum state $\rho$ and $\sigma$. Since $\rho$ and $\sigma$ are density matrices, 
$|\boldsymbol{x}_i\rangle$ and $|\boldsymbol{x}_i\rangle_{adv}$ should also be transformed from state vectors representation to their density matrix representation $|\boldsymbol{x}_i\rangle\langle\boldsymbol{x}_i|$ and $|\boldsymbol{x}_i\rangle_{adv}\langle\boldsymbol{x}_i|_{adv}$, respectively. 
$\langle\boldsymbol{x}_i|$ is the vector dual to $|\boldsymbol{x}_i\rangle$.
The fidelity is defined as:
\begin{equation}
	 F(\rho, \sigma) = \text{tr} \sqrt{\sqrt{\rho}\sigma\sqrt{\rho}} = |\langle\psi_\rho|\psi_\sigma\rangle|^2.
\end{equation}
The minimum overlap area of the probability distributions corresponding to two quantum states obtained by an optimal positive-operator valued measure (POVM) can intuitively reflect the fidelity between them \cite{Fidelity}. 
Figure \ref{fig:4} shows the visualization of fidelity between two quantum states based on POVM measurements. The greater the area, the higher the fidelity.
\end{definition}

\begin{definition} [Average trace distance, ATD] 
Similar to AFM, we define ATD as the average trace distance between the original inputs and their generated test samples. More precisely, the ATD is defined as: 
\begin{equation}
	ATD(\mathcal{T}) =\frac1{|\mathcal{T}|} \sum_{x_i \in \mathcal{T}} D(|\boldsymbol{x}_i\rangle\langle\boldsymbol{x}_i|,|\boldsymbol{x}_i\rangle_{adv}\langle\boldsymbol{x}_i|_{adv}),
\end{equation}
where $D(\rho, \sigma)$ is the trace distance between density operator $\rho$ and $\sigma$, which is calculated as follows:
\begin{equation}
	 D(\rho, \sigma) = \frac12\text{tr}|\rho-\sigma|.
\end{equation}
Quantum trace distance can be regarded as a generalization of classical trace distance.
It represents an achievable upper bound on the total variation distance between probability distributions arising from measuring two quantum states \cite{QAML12, QCQI}. 
In fact, during simulation calculations, 
the results of the quantum trace distance and the $L_1$ distance are often very close to each other.
\end{definition}

\section{Evaluation}

We implemented QuanTest\footnote{https://github.com/am0x00/QuanTest} using the deep learning framework PaddlePaddle \cite{ma2019paddlepaddle}, 
which includes the quantum machine learning framework Paddle Quantum 2.4.0\footnote{https://github.com/PaddlePaddle/Quantum} as a development kit for constructing and training QNNs using numerical simulation. 
The implementation of our core logic consists of around 280 lines of Python code. Additionally, there are about 2600 lines of supporting code for evaluation purposes.
All development and evaluation experiments for QuanTest were conducted on a computer with an Intel(R) Xeon(R) Gold 6146 CPU @ 3.20GHz processor with 12 cores, 256GB of memory, an NVIDIA GeForce RTX 2080 Ti GPU, and Ubuntu 18.04 as the host OS.

\subsection {Research Questions}

We summarize research questions to show QuanTest's performance and effectiveness, which are listed as follows.
\begin{itemize}
    \item {\textbf{RQ1}: How effective is QuanTest in capturing the erroneous behaviors of QNN?}
    \item {\textbf{RQ2}: How is the quality of quantum adversarial examples generated by QuanTest?}
    \item {\textbf{RQ3}: How is the effectiveness of entanglement guidance for QuanTest?}
    \item {\textbf{RQ4}: How adversarial examples generated by QuanTest affect the sampling cost of the QNN system?}
    \item {\textbf{RQ5}: How does QuanTest perform compared to black-box baseline?}
\end{itemize}

To demonstrate the efficacy of QuanTest in capturing erroneous behaviors of QNN systems,
we conducted tests using QuanTest on 9 representative mainstream QNN models. 
Subsequently, an analysis was performed on the generation of test samples by QuanTest across varying numbers of iterations. 
Finally, we evaluated the performance of QuanTest in quantum systems with varying dimensions. 
To answer RQ2, we analyze the fidelity measure and trace distance of the test samples generated by QuanTest, without imposing any similarity constraints.
Additionally, we explored the performance of the test samples in improving model performance after retraining.
To address RQ3, we conducted control experiments pertaining to entanglement-guided settings and analyzed the performance of QuanTest under the guidance of QEA with different values of $k$.
For RQ4, we model the measurement of the predicted probability qubit as a sampling problem in a Bernoulli distribution, based on which we explored the impact of the adversarial examples generated by QuanTest on the QNN's prediction under different sampling times.
Finally, we compare the performance of QuanTest with state-of-the-art black-box baseline, and discuss the usage scenarios of QuanTest.

\subsection {Experiment Setup}

As depicted in Table~\ref{table:1},
we adopt two widely utilized public datasets and then set three classification tasks. 
For each classification task, we evaluate QuanTest on three QNN models with distinct architectures (i.e., a total of nine QNN systems).
This allows us to cover various popular QNN variants and ensures the versatility of the experimental results.
Each real-world dataset is encoded into quantum state data using the same downscaling method and quantum state encoding strategy. 
All QNN models used for evaluation are trained by us using encoded quantum state data.
\subsubsection {Dataset}

\begin{table*}

\centering
  \caption{Details of Datasets and QNN Models.}
  \label{table:1}
  \resizebox{\linewidth}{!}{
  \begin{tabular}{ccccccc}
    \toprule
    Dataset & Description & Classification Task & \makecell{QNN\\ Model} & \makecell{\# Quantum\\ Gates} & \makecell{\# Trainable\\ Parameters} & Acc. (\%)\\
    \midrule
    \multirow{6}*{MNIST}  & \multirow{6}{14em}{28$\times$28 handwritten digit images, transformed into 16$\times$16 through downscaling, and embedded into 8-qubit states using amplitude encoding.} & \multirow{3}{8.3em}{Binary classification for images of digit 3 and 6} & CCQC & 200 & 200 & 98.68 \\ & & & QCL & 160 & 120 & 97.97 \\ & & & QCNN & 134 & 169 & 98.48 \\ \cmidrule(lr){3-7}& & \multirow{3}{8.6em}{Ternary classification for images of digit 0, 1 and 2} & CCQC & 200 & 200 & 88.63 \\ & & & QCL & 160 & 120 & 86.37 \\ & & & QCNN & 134 & 169 & 89.27  \\
    \midrule
    \multirow{3}*{\makecell{Fashion-\\MNIST}} & \multirow{3}{14.6em}{28$\times$28 grayscale images, undergoing the same downscaling method and encoding processing as MNIST.} & \multirow{3}{8.5em}{Binary classification for images of trouser and coat}  & CCQC & 200 & 200 & 95.85\\& & & QCL & 160 & 120 & 96.75\\& & & QCNN & 134 & 169 & 95.05\\
    
  \bottomrule
\end{tabular}
}
\end{table*}

The MNIST \cite{lecun2010mnist} handwritten digit database is a widely used classic dataset in the field of deep learning. 
It has become a standard benchmark for classification tasks. 
The dataset consists of 60,000 training examples and 10,000 testing examples. 
Each digit is stored as a gray-scale image with a size of 28$\times$28 pixels.
The Fashion-MNIST \cite{DBLP:journals/corr/abs-1708-07747} is a dataset of Zalando's article images containing 60,000 gray-scale fashion images as a training set and 10,000 images for testing.

We slightly reduced the size of images in both datasets from 28$\times$28 pixels to 16$\times$16 pixels so that we can simulate the QNN system with moderate classical computing resources. 
Subsequently, we established three classification tasks based on these two datasets, including binary classification of digits 3 and 6 in MNIST (denoted as MNIST[3,6]), ternary classification of digits 0, 1, and 2 in MNIST (denoted as MNIST[0,1,2]), and binary classification of trousers and coats in Fashion-MNIST (denoted as Fashion[1,4]). 
The selection of datasets and corresponding classification tasks is determined after a comprehensive consideration of the current limitations on the available qubits and the widely used setup of classification tasks in the field of quantum machine learning.

\subsubsection {QNN Models}

Quantum Circuit Learning (QCL) \cite{mitarai2018quantum} is a representative hybrid classical-quantum framework. 
It employs a quantum neural network composed of a nonlinear quantum encoding circuit for input data and a low-depth quantum variational circuit capable of learning tasks through iterative parameter optimization to accomplish machine learning tasks such as high-dimensional regression or classification.

Circuit-centric quantum classifiers (CCQC) \cite{schuld2020circuit} is a low-depth hybrid quantum neural network framework designed for supervised learning, 
which leverages the entangling properties of quantum circuits to capture the correlations within data and achieve classification tasks.

Quantum Convolutional Neural Network (QCNN) \cite{cong2019quantum} is a promising quantum neural network architecture inspired by classical convolutional neural networks.
It primarily consists of convolutional layers, pooling layers, and fully connected layers implemented using parameterized quantum circuits. 
QCNN is considered a promising architecture due to the absence of barren plateaus \cite{pesah2021absence}.

\subsubsection {Data Encoding}
Quantum neural networks employ quantum states as input. 
Therefore, when they are utilized to analyze classical data, 
we need first to consider an appropriate quantum state representation for classical world data in the quantum system. 
This is achieved through quantum state encoding, 
which implements a feature map from classical bits in an $N$-dimensional Euclidean space to $n$ qubits residing in a $2^n$-dimensional Hilbert space: $\mathbb{R}^N \to \mathbb{C}^{2^n}$, 
completes the embedding of classical data into quantum states.
There are several common strategies for quantum state encoding \cite{larose2020robust, schuld2018supervised, schuld2021supervised, PowerdatainQML}. 
To circumvent the challenges faced by traditional computers in numerically simulating a large number of qubits,
in this work, we exclusively focus on processing the adopted dataset through amplitude encoding.

Amplitude encoding maps the feature information of classical data to the amplitudes of quantum states. 
For the input classical data ${\bf{x}} \in \mathbb{R}^N$, 
its vector elements are encoded and mapped to the amplitudes of $\lceil \log_{2}N \rceil$ qubits: $|{\bf{x}}\rangle = \sum_{i=1}^N x_i |i\rangle$, 
where $\{|i\rangle\}$ represents a set of computational basis states in the Hilbert space. 
Since the feature information of classical data is represented as the amplitudes of a quantum state, 
the classical input must satisfy the normalization condition: $\|{\bf{x}} \|^2=\sum_i|x_i|^2=1$. 
 
We require the use of 256 amplitudes of the 8-qubit system to encode this data information, 
as each image in the MNIST and Fashion-MNIST datasets is two dimensional and contains 16$\times$16 pixels.
Note that QuanTest also possesses universality and extensibility for other encoding strategies.
 
\subsubsection {Test Samples Generation}
We use adversarial attack strategies FGSM \cite{FGSM} and DLFuzz \cite{DLFuzz} under the premise of an entanglement-guided approach to search for erroneous behaviors and generate quantum adversarial examples,
for testing QNN models on the MNIST and Fashion-MNIST datasets encoded as quantum states. 
We have adapted these adversarial attack strategies with quantum-based adjustments to enable them to process quantum state inputs and the information extracted within QNN systems.
Essentially, both of these strategies rely on a gradient-based manner to solve the optimization problem.
These adversarial examples have imperceptible perturbations that are difficult to detect, and demonstrate considerable effectiveness in revealing robustness issues in QNN systems.

In addition, we have prepared random noise samples generated by applying random coherent noise to original samples for subsequent comparative evaluation. 
We realized coherent noise by constructing a universal quantum rotation gate with Gaussian noise perturbation on the rotation parameters. 
This allows us to apply it as a random perturbation operator $U_{\Delta}(\sigma)$ on quantum original samples in the form of a quantum circuit, where $\sigma$ represents the strength of the noise, reflecting the magnitude of the standard deviation in the normal distribution.
Note that all samples need to be able to flip the model's predictions and pass quality evaluation to be output as test samples.

\subsubsection {Evaluation Metrics}

To assess the effectiveness of QuanTest, we define the test sample generation rate (Gen\_Rate), 
which represents the ratio of the original inputs the method has successfully produced quantum test samples:
\begin{equation}
	\text{Gen\_Rate} =\frac{|S_{adv}|}{|inputs|},
\end{equation}
where $S_{adv}$ represents the generated adversarial examples that have successfully passed quality evaluation (Algorithm \ref{algorithm1} line 1, 15-18), and $inputs$ represents the original input set (Algorithm \ref{algorithm1} line 2).
For the quality of the generated test samples, we introduce the AFM and ATD defined in Section 4.2.3 to assist us in evaluation.

\subsection{Capabilities of Erroneous Behavior Detection}

To address the \textbf{RQ1}, we first compared our QuanTest method with a strategy based on random coherent noise on multiple tested models. 
We randomly selected 800 original samples for each class category in the binary and ternary classification tasks of the MNIST and the binary classification task of Fashion-MNIST.
These samples were used to generate adversarial and random noise samples. 
For the generation of quantum adversarial examples, 
we set all hyperparameters in the adversarial attack strategy to 1.
We utilize QEA with $k=1$ as guidance, 
and the weight selected for joint optimization is set to $w=1$.
For the generation of random noise samples, 
We set the noise strength in the random perturbation operator as $\sigma=0.02$.
Then, we choose a high standard for evaluating similarity with fidelity measure > 90\% or trace distance < 0.45 to determine the qualification of the test samples, 
given the extensive Hilbert space dimensions of the QNN system.
We ran the experiments three times on each tested model and took the average, to mitigate potential uncertainties in the results. 
In the absence of additional declarations, other evaluations default to the same predefined configuration.

Table~\ref{table:2} shows the overall results obtained with QuanTest and random coherent noise,
where Gen\_Rate represents the test sample generation rate.
The best performance is typeset in bold.
Test samples generated by each strategy exhibit comparable similarities, as restrictions are imposed on the similarity metrics.
The maximum difference for AFM does not exceed 3.89\%, and for ATD, it does not exceed 0.14. 
This allows the evaluation of the test sample generation rates for each strategy with minimal interference. 
It can be observed that the test sample generation rate of coherent noise is generally low, reaching a maximum of only 4.58\%. 
In contrast, QuanTest achieves a maximum test sample generation rate of 84.59\% using the FGSM attack strategy, while the use of the DLFuzz attack strategy can result in a success rate of up to 96.87\%. 
The test sample generation rate of the DLFuzz strategy is 9.53\% to 26.43\% higher than that of the FGSM strategy across various models.

\begin{table}
  \caption{Capability of QuanTest to capture erroneous behaviors based on different adversarial attack strategies.}
  \label{table:2}
\resizebox{\columnwidth}{!}{
\renewcommand\arraystretch{1.15}
  \begin{tabular}{ccccccccccc}
    \hline
        \multicolumn{1}{c|}{\multirow{2}{*}{Data}}             & \multicolumn{1}{c|}{\multirow{2}{*}{Model}} & \multicolumn{3}{c|}{Gen\_Rate(\%)}                                                                                                                                                                             & \multicolumn{3}{c|}{AFM(\%)}                                                                                                                                                                                  & \multicolumn{3}{c}{ATD}                                                                                                                                                                                  \\ \cline{3-11} 
    \multicolumn{1}{c|}{}                                  & \multicolumn{1}{c|}{}                       & \multicolumn{1}{c}{Random} & \multicolumn{1}{c}{\begin{tabular}[c]{@{}c@{}}QuanTest\\      (FGSM-based)\end{tabular}} & \multicolumn{1}{c|}{\begin{tabular}[c]{@{}c@{}}QuanTest\\      (DLFuzz-based)\end{tabular}} & \multicolumn{1}{c}{Random} & \multicolumn{1}{c}{\begin{tabular}[c]{@{}c@{}}QuanTest\\      (FGSM-based)\end{tabular}} & \multicolumn{1}{c|}{\begin{tabular}[c]{@{}c@{}}QuanTest\\      (DLFuzz-based)\end{tabular}} & \multicolumn{1}{c}{Random} & \multicolumn{1}{c}{\begin{tabular}[c]{@{}c@{}}QuanTest\\      (FGSM-based)\end{tabular}} & \multicolumn{1}{c}{\begin{tabular}[c]{@{}c@{}}QuanTest\\      (DLFuzz-based)\end{tabular}} \\ \hline
    \multicolumn{1}{c|}{\multirow{3}{*}{Fashion{[}1,4{]}}} & \multicolumn{1}{c|}{CCQC}                  & 0.62&73.57& \multicolumn{1}{c|}{\bf87.79}                                                                                                                                                                           &     \bf97.58&  93.65        & \multicolumn{1}{c|}{93.52}                                                                                                                                                                                &  \bf0.2143  &    0.3376    & \multicolumn{1}{c}{0.3428}                                                                                                                                                                                     \\ \cline{2-11} 
    \multicolumn{1}{c|}{}                                  & \multicolumn{1}{c|}{QCL}                  & 0.83&75.84 & \multicolumn{1}{c|}{\bf96.48}                                                                                                                                                                           &    \bf94.29 &    93.60         & \multicolumn{1}{c|}{93.90}                                                                                                                                                                              &  \bf0.3138   &   0.3394       & \multicolumn{1}{c}{0.3327}                                                                                                                                                                                     \\ \cline{2-11} 
    \multicolumn{1}{c|}{}                                  & \multicolumn{1}{c|}{QCNN}                 & 2.49&79.94 & \multicolumn{1}{c|}{\bf92.73}                                                                                                                                                                            &    \bf 95.80 &   95.09         & \multicolumn{1}{c|}{95.07}                                                                                                                                                                               &  \bf0.2669   &   0.2924      & \multicolumn{1}{c}{0.2947}                                                                                                                                                                                     \\ \hline
    \multicolumn{1}{c|}{\multirow{3}{*}{MNIST{[}0,1,2{]}}}   & \multicolumn{1}{c|}{CCQC}                  & 3.15&44.52& \multicolumn{1}{c|}{\bf70.95}                                                                                                                                                                          &  \bf 95.65   &    95.28          & \multicolumn{1}{c|}{95.06}                                                                                                                                                                             &  0.2792   &    \bf0.2775       & \multicolumn{1}{c}{0.2852}                                                                                                                                                                                     \\ \cline{2-11} 
    \multicolumn{1}{c|}{}                                  & \multicolumn{1}{c|}{QCL}                   & 3.74&56.17& \multicolumn{1}{c|}{\bf71.52}                                                                                                                                                                            &   95.40  &    95.55        & \multicolumn{1}{c|}{\bf96.26}                                                                                                                                                                               &  0.2843   &   0.2677      & \multicolumn{1}{c}{\bf0.2467}                                                                                                                                                                                     \\ \cline{2-11} 
    \multicolumn{1}{c|}{}                                  & \multicolumn{1}{c|}{QCNN}                 &  4.58&69.71& \multicolumn{1}{c|}{\bf79.24}                                                                                                                                                                            &   95.36  &    97.22        & \multicolumn{1}{c|}{\bf97.41}                                                                                                                                                                               &  0.2827   &    0.2058     & \multicolumn{1}{c}{\bf0.1984}                                                                                                                                                                                     \\ \hline
    \multicolumn{1}{c|}{\multirow{3}{*}{MNIST{[}3,6{]}}} & \multicolumn{1}{c|}{CCQC}                &  1.02&70.42 & \multicolumn{1}{c|}{\bf88.44}                                                                                                                                                                              &   93.13  &    94.35      & \multicolumn{1}{c|}{\bf94.54}                                                                                                                                                                                  &  0.3611   &  0.3153    & \multicolumn{1}{c}{\bf0.3117}                                                                                                                                                                                     \\ \cline{2-11} 
    \multicolumn{1}{c|}{}                                  & \multicolumn{1}{c|}{QCL}                  &  2.26&59.00& \multicolumn{1}{c|}{\bf91.70}                                                                                                                                                                            &   95.24  &    94.76        & \multicolumn{1}{c|}{\bf95.44}                                                                                                                                                                                 &  \bf0.2800   &  0.3540     & \multicolumn{1}{c}{0.2811}                                                                                                                                                                                     \\ \cline{2-11} 
    \multicolumn{1}{c|}{}                                  & \multicolumn{1}{c|}{QCNN}                 &  0.82&84.59& \multicolumn{1}{c|}{\bf96.87}                                                                                                                                                                            &   93.85  &     94.95       & \multicolumn{1}{c|}{\bf95.29}                                                                                                                                                                                 &  0.3398   &   0.2963    & \multicolumn{1}{c}{\bf0.2881}                                                                                                                                                                                     \\ \hline
    \end{tabular}}
\end{table}

Therefore, it can be concluded that QuanTest's ability to capture erroneous behaviors is significantly higher under similar levels of similarity compared to the random coherent noise used as the baseline, 
and the adversarial attack strategies based on DLFuzz often generate more test samples.
Note that the performance of QNN models in the presence of coherent noise also demonstrates its significant robustness to quantum stochastic noise itself.
Models with better pre-training (The accuracy is shown in Table~\ref{table:1}) exhibit stronger resistance to coherent noise.
In addition, we observe a certain negative correlation between the model's resistance to coherent noise and its performance in QuanTest (Pearson correlation coefficient of -0.74 for the test sample generation rate). 
This may imply that QuanTest is more capable of identifying deeper, atypical security issues in QNN systems.

\begin{figure}[tb]
  \centering
  \includegraphics[width=\linewidth]{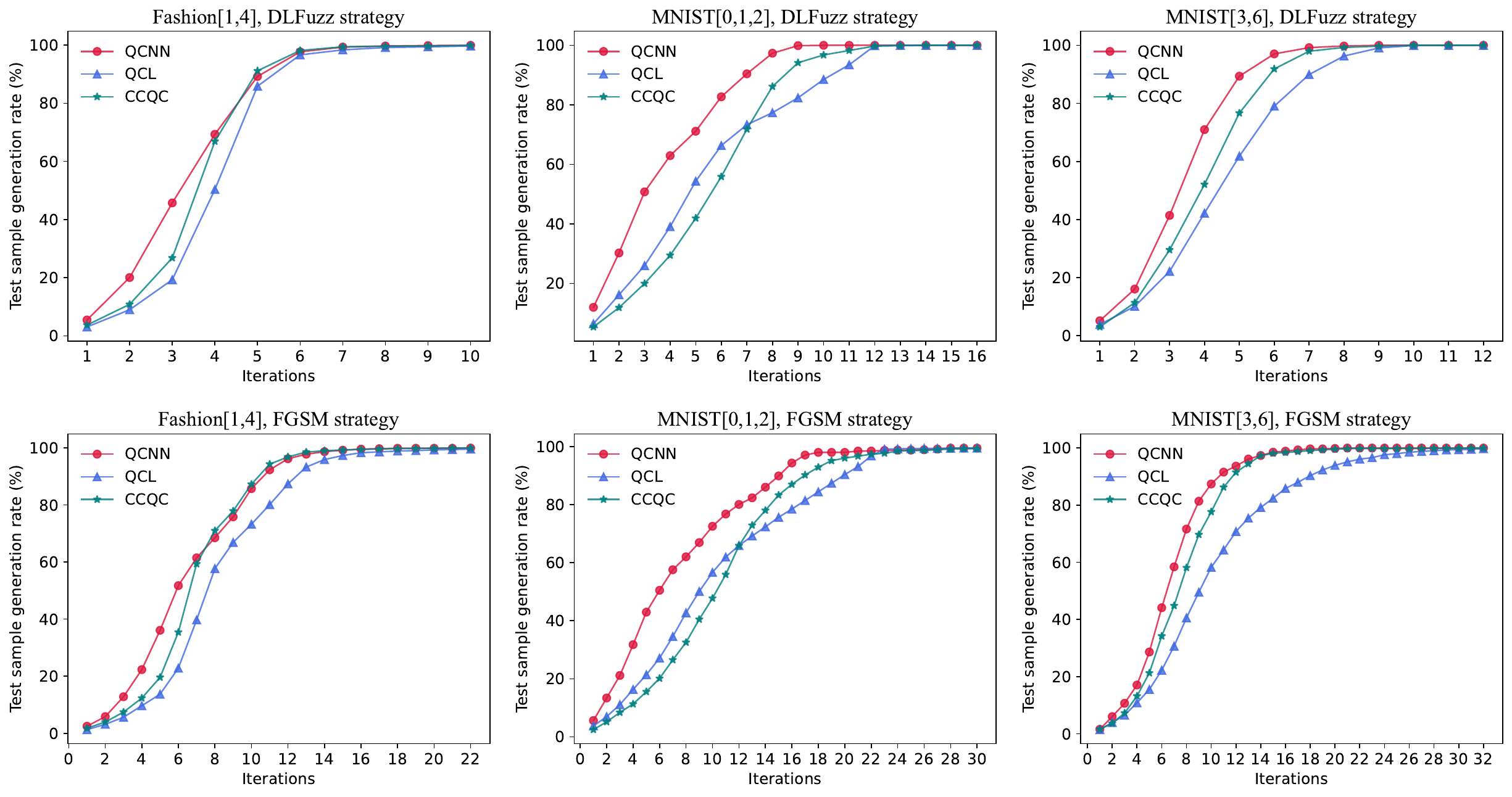}
   \caption{Test sample generation rate of QuanTest at different iterations.}
  \Description{Test sample generation rate of QuanTest at different iterations.}
  \label{fig:5}
\end{figure}

Subsequently, we calculated the test sample generation rate under different iteration counts using the entanglement-guided DLFuzz strategy when no similarity constraints were imposed (i.e., Gen\_Rate=100\%), as shown in Figure \ref{fig:5}.
The results indicate that QuanTest based on the DLFuzz strategy is capable of generating the majority of adversarial examples after 5-10 iterations.
In contrast, QuanTest based on the FGSM strategy requires slightly more iterations. 
This may be related to the effectiveness of the adversarial attack strategy.
In addition, the speed of test sample generation for QCNN is generally faster than other models, which may be related to the quality of the model construction.

We also investigate the performance of QuanTest in capturing erroneous behaviors when scaling to higher-dimensional quantum systems.
We used 784 amplitudes from the 10-qubit system to prepare quantum states for images of size 28$\times$28 pixels in the dataset, with the last 240 bits of the state vector padded with zeros to ensure a length of $2^{10}$. 
We extended the pre-trained QNN model accordingly, increasing the number of quantum gates and trainable parameters by 0.25 times due to the increase in the number of qubits. 
The training was terminated when the model achieved performance similar to the QNN model corresponding to 16$\times$16 pixel images in the classification. 
This was done to ensure fairness in the comparative experiments.
For each classification category in the classification task, we randomly selected 500 samples of size 16$\times$16 pixels and 28$\times$28 pixels. fidelity measure > 95\% or trace distance < 0.30 were chosen as the criteria for adversarial example detection.

\begin{table}
  \caption{Performance of QuanTest based on the DLFuzz strategy in capturing erroneous behaviors in quantum systems of different dimensions.}
  \label{table:3}
  \resizebox{0.5\columnwidth}{!}{
\begin{tabular}{ccccc}
\toprule
\multirow{2}{*}{Performance} & \multirow{2}{*}{Model} & \multirow{2}{*}{Data}  & \multicolumn{2}{c}{Dimension} \\\cmidrule(l{0.5em}r){4-5}
                                     &                        &                   &  $16 \times 16$  & $28 \times 28$ \\
\midrule
\multirow{9}{*}{Gen\_Rate(\%)}      & \multirow{3}{*}{CCQC}           &  Fashion[1,4]       & 26.87    & \bf84.90      \\
                                     &                                  &  MNIST[0,1,2]       & 30.59    & {\bf82.91}    \\
                                     
                                     &                                  &  MNIST[3,6]        & 47.75    & \bf87.50   \\

                                     \cmidrule(lr){2-5}
                                     & \multirow{3}{*}{QCL}   & Fashion[1,4]                        & 37.78    & \bf70.14    \\
                                     &                        &  MNIST[0,1,2]         & 46.51    & {\bf59.82} \\
                                     
                                     &                        &  MNIST[3,6]                      & 65.30    & \bf78.72     \\
                                     \cmidrule(lr){2-5}
                                     & \multirow{3}{*}{QCNN}   & Fashion[1,4]                        & 39.20    & \bf89.29   \\
                                     &                        &  MNIST[0,1,2]         & 53.55    & {\bf97.82} \\
                                     
                                     &                        &  MNIST[3,6]                      & 53.78    & \bf94.33     \\                                     
\midrule
\multirow{9}{*}{AFM(\%)}        & \multirow{3}{*}{CCQC}  &  Fashion[1,4]                        & \bf96.94    & 96.89     \\
                                     &                        &  MNIST[0,1,2]         & 96.92    & {\bf98.01}  \\
                                     
                                     &                        &   MNIST[3,6]                      & 97.26    & \bf98.58     \\
                                     \cmidrule(lr){2-5}
                                     & \multirow{3}{*}{QCL}   & Fashion[1,4]                        & \bf97.79    &\bf 97.79    \\
                                     &                        &  MNIST[0,1,2]         & 97.96    & {\bf98.40} \\
                                     
                                     &                        &  MNIST[3,6]                      & 98.45    & \bf98.52     \\                                     
                                     \cmidrule(lr){2-5}
                                     & \multirow{3}{*}{QCNN}  &  Fashion[1,4]                       & 97.18    & \bf97.63     \\
                                     &                        &  MNIST[0,1,2]         & 97.37    & {\bf98.13}  \\
                                     
                                     &                        &   MNIST[3,6]                     & 97.22    & \bf97.75      \\
\midrule
\multirow{9}{*}{ATD}                & \multirow{3}{*}{CCQC}  &  Fashion[1,4]                        & \bf0.2344    & 0.2367     \\
                                     &                        &  MNIST[0,1,2]         & 0.2349    & {\bf0.1914}  \\
                                     
                                     &                        &   MNIST[3,6]                      & 0.2227    &\bf 0.1622     \\
                                     \cmidrule(lr){2-5}
                                     & \multirow{3}{*}{QCL}   &  Fashion[1,4]                       & \bf0.1922    &0.1950    \\
                                     &                        & MNIST[0,1,2]        & 0.1835    & {\bf0.1629}  \\
                                     
                                     &                        &   MNIST[3,6]                     & 0.1606    & \bf0.1560      \\
                                     \cmidrule(lr){2-5}
                                     & \multirow{3}{*}{QCNN}  &  Fashion[1,4]                       & 0.2269    & \bf0.2057     \\
                                     &                        &  MNIST[0,1,2]       & 0.2164    & {\bf0.1808}   \\
                                     
                                     &                        &   MNIST[3,6]                      & 0.2241    & \bf0.1998    \\
                                     
\bottomrule
\end{tabular}
}
\end{table}

Table~\ref{table:3} shows the QuanTest's performance using the DLFuzz strategy on samples of 16$\times$16 pixels and the corresponding samples of 28$\times$28 pixels.
The best result across each row is denoted in bold.
The results indicate that, under the same similarity constraints, QuanTest tends to generate more test samples for quantum systems with higher dimensions. 
This finding also partially confirms the theoretical proof by Liu and Wittek, demonstrating that a disturbance inversely proportional to the dimension is sufficient to cause a QNN system to induce a misclassification \cite{RQD}.

\begin{center}
\begin{tcolorbox}[colback=gray!10,
                  colframe=black,
                  width=\linewidth,
                  arc=1mm, auto outer arc,
                  boxrule=0.5pt,
                 ]
{\bf Answer to RQ1:} QuanTest is capable of effectively capturing erroneous behaviors in QNN systems. It can efficiently generate a sufficient number of test samples with slight perturbations and performs better in quantum systems with higher dimensions.
\end{tcolorbox}
\end{center}

\subsection {Quality of Generated Adversarial Examples}
\begin{figure*}[tp]
  \centering
  \includegraphics[width=\linewidth]{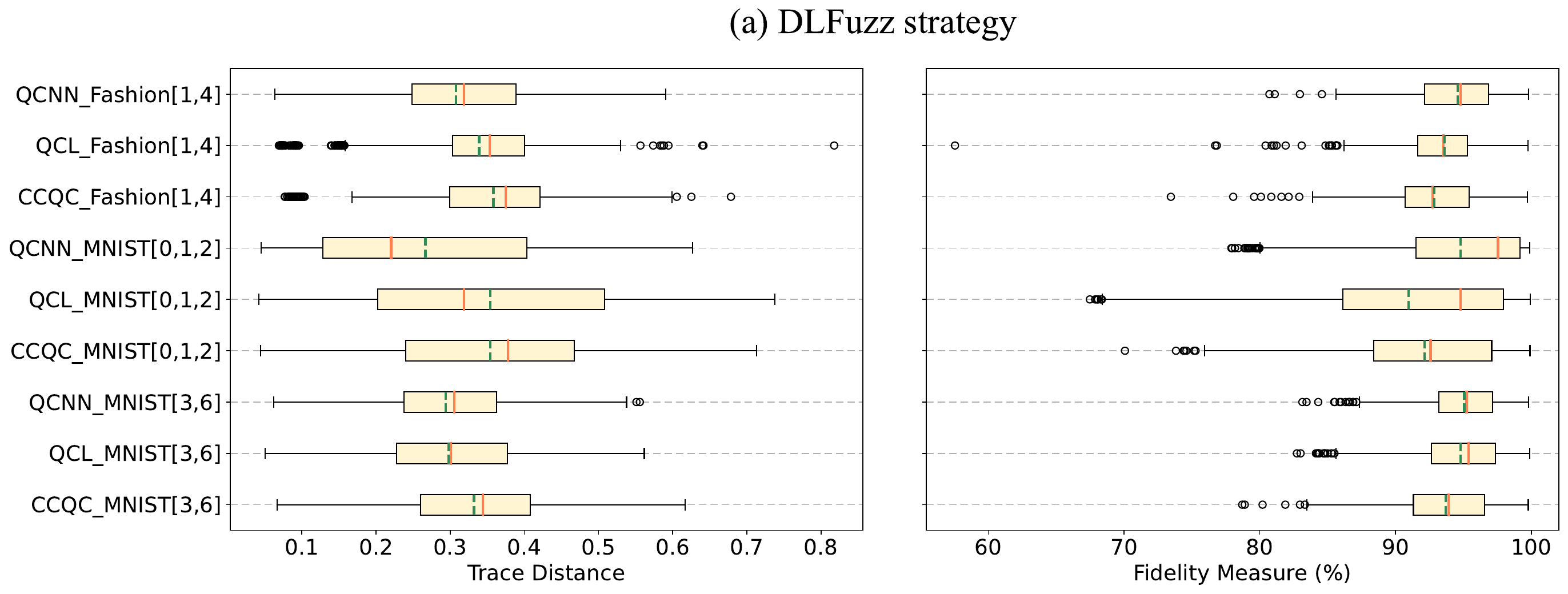}
  \includegraphics[width=\linewidth]{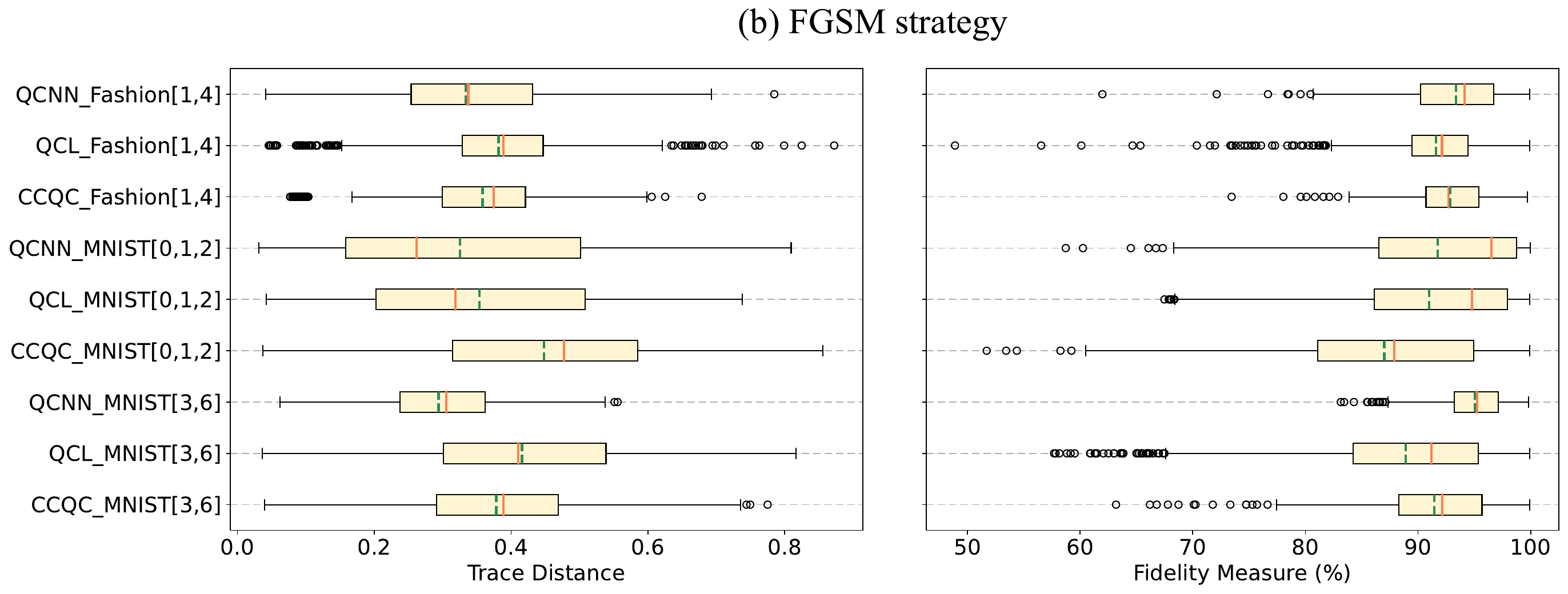}
  \caption{Value range of trace distance and fidelity measure of adversarial examples generated by entanglement-guided adversarial attack strategies without imposing similarity constraints.}
  \Description{Value range of trace distance and fidelity measure of adversarial examples generated by entanglement-guided adversarial attack strategies without imposing similarity constraints.}
  \label{fig:6}
\end{figure*}
The quality of test samples can be multifaceted. 
To address the \textbf{RQ2}, we measure the quality of adversarial examples generated by QuanTest by evaluating the similarity between adversarial examples and original samples, as well as their improvement in enhancing the performance of QNN systems after retraining.
We conducted a statistical analysis of the fidelity measure and trace distance between the quantum adversarial examples generated by QuanTest using different adversarial attack strategies and the quantum original samples. 
QuanTest generated 600 adversarial examples for each classification task without imposing any similarity constraints.
In other words, we assessed the quality of the adversarial examples generated by QuanTest under the condition of Gen\_Rate=100\%.
This condition setting enables our statistics to cover a wide range of scenarios, including extreme cases, thereby enhancing the comprehensiveness of the evaluation.

Figure \ref{fig:6} presents the value range of fidelity measure and trace distance obtained by different methods on each classification task respectively.
The solid coral line corresponds to the median, while the dashed green line signifies the mean (i.e., AFM and ATD).
We can observe that the fidelity measure of adversarial examples generated by QuanTest based on the DLFuzz strategy is generally higher than 85\%, and the trace distance is generally lower than 0.6.
The performance of using the FGSM strategy is slightly less favorable than the DLFuzz strategy,
but the fidelity is generally above 80\%, and the trace distance is generally below 0.8. 
Both strategies exhibit similar statistical patterns overall.
The differences in the quality of test samples generated by different models based on the two strategies are also consistent with the differences in the iteration counts shown in Figure \ref{fig:5}.

In fact, 80\% is still a relatively high fidelity, given that the Hilbert space dimension of the QNN model is already very large.
Therefore, whether using the DLFuzz or FGSM strategy, the adversarial examples generated under the guidance of QEA exhibit remarkable performance on AFM and ATD. 
This implies that the majority of test samples generated by QuanTest can trigger erroneous behaviors in the QNN system by applying imperceptible perturbations. 
This also allows us to generate test samples using QEA guidance without needing to overly consider the additional distortion cost produced by the joint optimization process. 

We subsequently merged the generated test samples from Section 4.3 into the corresponding training sets of classification tasks, and retrained the QNN models with the same settings to investigate whether the test samples generated by QuanTest can be used for retraining to further improve model performance.
In addition to using test samples generated by single strategies to constitute enhanced datasets, we also merged the test samples generated by QuanTest based on two adversarial attack strategies, DLFuzz and FGSM, with the original dataset to form a new enhanced dataset, aiming to explore the performance of mixed test samples generated by multiple strategies.

\begin{figure}[tbp]
  \centering
  \includegraphics[width=\linewidth]{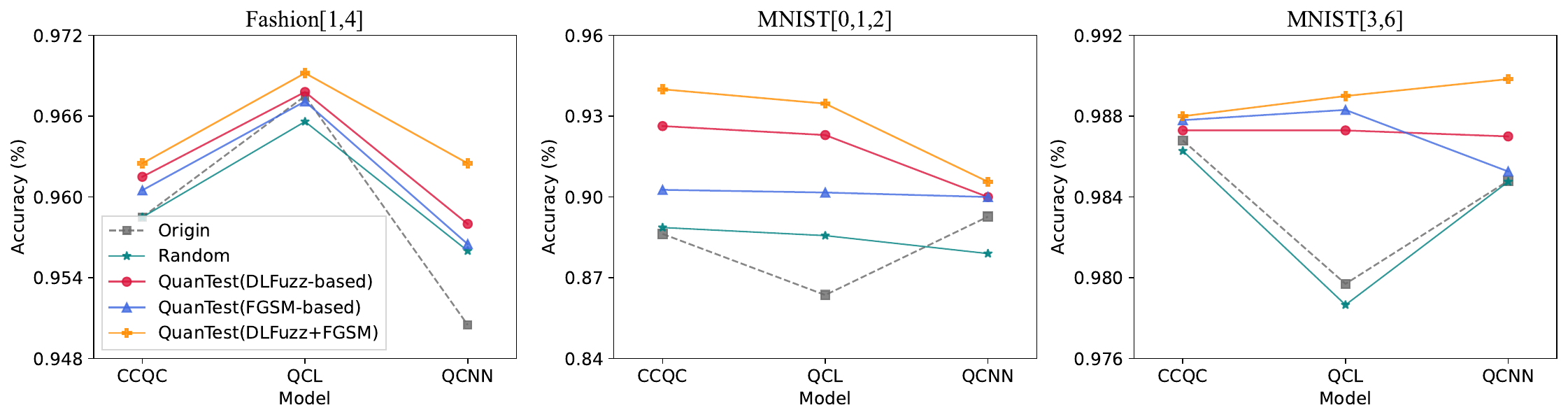}
  \caption{Accuracy of three QNN models for the MNIST binary classification task before and after retraining.}
  \Description{The accuracy of three QNN models for the MNIST binary classification task before and after retraining.}
  \label{fig:5.5}
\end{figure}

Figure \ref{fig:5.5} shows the accuracy of the model after retraining on different enhanced datasets. 
Here the data are averaged over three times of the same retraining process to mitigate the impact arising from uncertainty.
It can be seen that the test samples generated by QuanTest can enhance the training effect of the model by adding more corner cases. 
Models with lower original accuracy often gain more improvement after retraining, especially the enhanced training set composed of mixed test samples, which can bring up to a maximum of 7.1\% improvement.
This means that adversarial examples generated based on different attack strategies can discover different types of erroneous behaviors, thereby improving the robustness of QNN systems.

\begin{center}
\begin{tcolorbox}[colback=gray!10,
                  colframe=black,
                  width=\linewidth,
                  arc=1mm, auto outer arc,
                  boxrule=0.5pt,
                 ]
{\bf Answer to RQ2:} The adversarial examples generated by QuanTest exhibit high quality and perform well in terms of both fidelity measure and trace distance. 
They can be utilized for retraining to enhance the robustness of QNN systems.
\end{tcolorbox}
\end{center}

\subsection {Effectiveness of Entanglement Guidance}

To address the \textbf{RQ3}, we first evaluated the performance of the tested model under two settings: with entanglement guidance and without entanglement guidance (i.e., pure adversarial attacks).
In the absence of a similarity threshold, QuanTest randomly selected 600 original samples for each classification task to generate test samples, subjecting each quantum original sample to 5 iterations.
We repeat each configuration 5 times and report the average results to mitigate randomness in the evaluation process.

\begin{table}
  \caption{Performance of QuanTest with and without entanglement guidance.}
  \label{table:4}
  \resizebox{0.7\columnwidth}{!}{
\begin{tabular}{@{}p{5.4em}<{\centering}p{3.2em}<{\centering}p{3.4em}<{\centering}p{3.2em}<{\centering}p{5.0em}<{\centering}p{3.2em}<{\centering}p{4.5em}<{\centering}@{}}
\toprule
\multirow{2}{*}{Data} & \multirow{2}{*}{Model} & \multirow{2}{*}{Strategy}  & \multicolumn{4}{c}{Performance} \\\cmidrule(l{0.2em}r){4-7}
                                     &                        &                          &  {\makecell{QEA(\%)\\ ($k=1$)}}  & Gen\_Rate(\%)  & AFM(\%)    & ATD\\
\midrule
\multirow{12}{*}{Fashion[1,4]}          & \multirow{4}{*}{CCQC}  &  $\mathcal{D}$                          & 49.55    & 71.49    & \bf94.84    & \bf0.3061  \\
                                     &                        &  $\mathcal{D} + \mathcal{G}$          & \bf57.12    & {\bf91.17}   & 93.42    & 0.3452   \\
                                     \cmidrule(lr){3-7}
                                     &                        &  $\mathcal{F}$                          & 51.39    & 11.78 & {\bf99.00}  & {\bf0.1305}  \\
                                     &                        &  $\mathcal{F} + \mathcal{G}$          & {\bf57.88}    & \bf19.60  & 97.41    & 0.2117   \\                                     

                                     \cmidrule(lr){2-7}
                                     & \multirow{4}{*}{QCL}   &  $\mathcal{D}$                      & 54.10    & 70.00   & \bf96.15    & \bf0.2587   \\
                                     &                        &  $\mathcal{D} + \mathcal{G}$         & \bf61.45    & {\bf85.86}  & 94.33    & 0.3212  \\
                                     \cmidrule(lr){3-7}
                                     &                        &  $\mathcal{F}$                      & 55.44    & 10.88   & {\bf99.21}  & {\bf0.1170}   \\
                                     &                        &  $\mathcal{F} + \mathcal{G}$         & {\bf62.12} & \bf13.69  & 98.04    & 0.1834  \\
                                     
                                       \cmidrule(lr){2-7}
                                     & \multirow{4}{*}{QCNN}   &  $\mathcal{D}$                      & 38.08    & 75.31   & \bf97.07    & \bf0.2225   \\
                                     &                        &  $\mathcal{D} + \mathcal{G}$         & \bf44.84    & {\bf89.22}  & 95.28    & 0.2890  \\
                                     \cmidrule(lr){3-7}
                                     &                        &  $\mathcal{F}$                      & 38.19    & 21.25   & {\bf99.06}  & {\bf0.1272}   \\
                                     &                        &  $\mathcal{F} + \mathcal{G}$         & {\bf44.93} & \bf36.08  & 97.61    & 0.2049  \\                                   
\midrule
\multirow{12}{*}{MNIST[0,1,2]}        & \multirow{4}{*}{CCQC}  &  $\mathcal{D}$                      & 16.25    & 33.11    & \bf98.98    & \bf0.1247  \\
                                     &                        &  $\mathcal{D} + \mathcal{G}$        & \bf24.53    & {\bf41.90}  & 97.32    & 0.2093 \\
                                     \cmidrule(lr){3-7}
                                     &                        &  $\mathcal{F}$                      & 17.05    & 14.22    & {\bf99.70}    & {\bf0.6983}  \\
                                     &                        &  $\mathcal{F} + \mathcal{G}$         & {\bf24.66}    & \bf15.55  & 98.78    & 0.1401  \\
                                      \cmidrule(lr){2-7}
                                      
                                     & \multirow{4}{*}{QCL}   &  $\mathcal{D}$                      & 19.93    & 48.60   & \bf98.85    & \bf0.1342   \\
                                     &                        &  $\mathcal{D} + \mathcal{G}$         & \bf28.00    & {\bf54.33}  & 97.37    & 0.2070  \\
                                     \cmidrule(lr){3-7}
                                     &                        &  $\mathcal{F}$                      & 19.80    & 18.28   & {\bf99.67}  & {\bf0.739}   \\
                                     &                        &  $\mathcal{F} + \mathcal{G}$         & {\bf26.82} & \bf21.39  & 98.81    & 0.1384  \\                                    
                                     \cmidrule(lr){2-7}
                                     & \multirow{4}{*}{QCNN}  &  $\mathcal{D}$                      & 18.09    & 69.34     & \bf98.66    & \bf0.1480 \\
                                     &                        &  $\mathcal{D} + \mathcal{G}$         & \bf22.89    & {\bf71.14}& 98.00    & 0.1794  \\
                                     \cmidrule(lr){3-7}
                                     &                        &  $\mathcal{F}$                      & 18.19    & 36.08     & {\bf99.60}  & {\bf0.8282} \\
                                     &                        &  $\mathcal{F} + \mathcal{G}$          & {\bf23.34} & \bf42.07    & 98.88    & 0.1364  \\                                     
\midrule
\multirow{12}{*}{\makecell{MNIST[3,6]}}   & \multirow{4}{*}{CCQC}  &  $\mathcal{D}$                      & 19.93    & 55.33    & \bf96.92    & \bf0.2312  \\
                                     &                        &  $\mathcal{D} + \mathcal{G}$        & \bf28.07    & {\bf76.65}  & 95.10    & 0.2958 \\
                                     \cmidrule(lr){3-7}
                                     &                        &  $\mathcal{F}$                      & 17.10    & 12.36    & {\bf99.12}    & {\bf0.1263}  \\
                                     &                        &  $\mathcal{F} + \mathcal{G}$         & {\bf26.10}    & \bf21.29  & 97.78    & 0.1973  \\
                                      \cmidrule(lr){2-7}
                                      
                                     & \multirow{4}{*}{QCL}   &  $\mathcal{D}$                      & 21.32    & 60.37    & \bf98.34    & \bf0.1676  \\
                                     &                        & $\mathcal{D} + \mathcal{G}$        & \bf31.79    & {\bf61.80} & 96.98    & 0.2314  \\
                                     \cmidrule(lr){3-7}
                                     &                        &  $\mathcal{F}$                      & 21.55    & 14.51    & {\bf99.59} & {\bf0.8431}  \\
                                     &                        &  $\mathcal{F} + \mathcal{G}$         & {\bf31.55} & \bf15.52   & 98.59    & 0.1553  \\                                     
                                     \cmidrule(lr){2-7}
                                     & \multirow{4}{*}{QCNN}  &  $\mathcal{D}$                      & 16.05    & 83.92    & \bf97.16    & \bf0.2213  \\
                                     &                        &  $\mathcal{D} + \mathcal{G}$        & \bf23.67    & {\bf89.37}  & 95.67    & 0.2773  \\
                                     \cmidrule(lr){3-7}
                                     &                        &  $\mathcal{F}$                      & 15.34    & 18.39   & {\bf99.16} & {\bf0.1212} \\
                                     &                        &  $\mathcal{F} + \mathcal{G}$        & {\bf23.34}& \bf28.63    & 98.06    & 0.1837  \\    
                                     
\bottomrule
\end{tabular}
} 
\end{table}

Table~\ref{table:4} presents the overall performance of the DLFuzz strategy (denoted as $\mathcal{D}$) and FGSM strategy (denoted as $\mathcal{F}$) in QuanTest under settings with and without entanglement guidance (denoted as $\mathcal{G}$).
We compare the performance of QuanTest when adopting the same adversarial attack strategy with or without the entanglement-guided settings, and highlight the better result in bold from the two scenarios.
In the case of the FGSM strategy, the entanglement-guided setting can achieve an improvement of 5.15\% to 10.00\% in QEA, and the test sample generation rate can be increased by up to 14.84\%. 
Similarly, in the case of the DLFuzz strategy, the entanglement-guided setting can achieve an improvement of 4.81\% to 10.47\% in QEA, with a maximum improvement of 21.32\% in the test sample generation rate.
In addition, settings with entanglement guidance also lead to lower similarity between the generated test samples and the original samples. 
Compared to settings without entanglement guidance, entanglement-guided settings reduce AFM by 0.65\% to 1.82\% and increase ATD by 0.29 to 0.85.
The results indicate that, based on the same adversarial attack strategy, entanglement-guided configurations often lead to higher QEA and generate more adversarial examples. 
Although entanglement guidance may introduce additional slight distortions, the marginal cost is entirely acceptable. 
Note that the increase in test sample generation rates brought about by entanglement-oriented settings cannot be achieved with the same small cost using random coherent noise.

\begin{figure}[tb]
  \centering
  \includegraphics[width=0.91\linewidth]{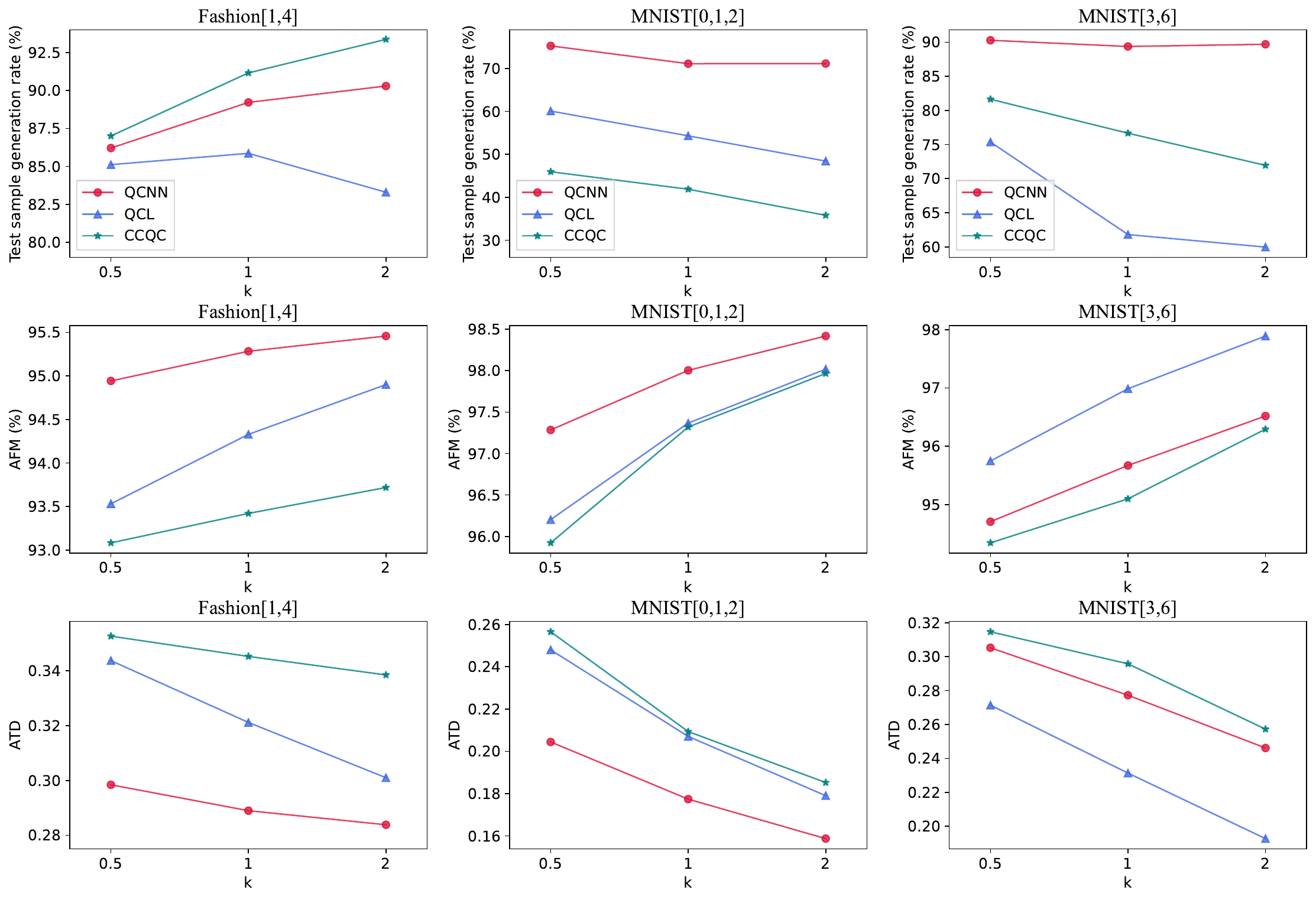}
  \caption{Performance of QuanTest based on the DLFuzz strategy under the guidance of QEA at different values of $k$.}
  \Description{.}
  \label{fig:9}
\end{figure}

In addition, we also assessed the performance of QuanTest based on the DLFuzz strategy under the guidance of QEA at different values of $k$.
The magnitude of the $k$ value can reflect the proportion of entanglement of the output quantum state in QEA.
Similarly, we do not set a similarity limit and set the iteration count for each original sample to 5. 
We use the QEA's variant $QEA(k) = \frac{2k}{k+1} \cdot Q(U_{\boldsymbol{\Theta}}|\boldsymbol{x}\rangle)-\frac{2}{k+1}Q(|\boldsymbol{x}\rangle)$ as the optimization objective in the joint optimization process to avoid significant differences in the perturbation amounts generated by QEA at different $k$ values, thereby avoiding interference with the evaluation. 
Figure \ref{fig:9} presents the overall results of the assessment.

For the MNIST dataset, the larger the $k$ value of QEA, the fewer test sample generation rates it can achieve, while the Fashion dataset does not exhibit this property. 
The influence of QEA with different $k$ values on the test sample generation rate further proves that entanglement guidance plays a role in generating adversarial examples.
Another apparent phenomenon is that, with controlled perturbation levels, the smaller the $k$ value of QEA, the lower the AFM of the generated test samples and the higher the ATD. 
This may be due to the concentration of entanglement measures of the output quantum state at higher values. 
This pattern also provides some reference for using QuanTest in the testing of QNN systems.
When aiming to generate test samples with the highest possible similarity to the original samples, it may be advisable to consider appropriately increasing the value of $k$ in QEA.
\begin{center}
\begin{tcolorbox}[colback=gray!10,
                  colframe=black,
                  width=\linewidth,
                  arc=1mm, auto outer arc,
                  boxrule=0.5pt,
                 ]
{\bf Answer to RQ3:} Entanglement guidance can effectively traverse the solution space of QNN systems, discovering more erroneous behaviors. This capability is related to the proportion of entanglement of the output quantum state in QEA.
\end{tcolorbox}
\end{center}

\section {Discussion}

\subsection{Effect of QuanTest on Sampling Cost}

In this section, we explore how adversarial examples generated by QuanTest affect the sampling cost of QNN system.
In order to read out the state of a qubit, we must measure it.
Unlike quantum state tomography, which reconstructs the complete quantum state through measurements, obtaining predictive estimates from the measurement results output by QNN models is a simpler problem.
Taking binary classification as an example, at least one qubit is needed to encode the prediction results of QNN model $\mathcal{M}$.
Assuming that the qubit is in state $|\mathcal{M}(x)\rangle=|1\rangle$ the prediction is 1, and if it is in the state $|\mathcal{M}(x)\rangle=|0\rangle$ the prediction is 0. 
Superposition can be interpreted as providing probability outputs regarding the uncertainty of the result. 
At this point, $\mathcal{M}(x)=p$ represents the probability that the model's prediction of $y=1$ for input $x$. 
In the case of error-free quantum algorithms, measuring a single predicted probability qubit of the QNN model $\mathcal{M}$ can evolve into a problem of sampling from a Bernoulli distribution.

\begin{figure*}[t]
  \centering
  \includegraphics[width=0.95\linewidth]{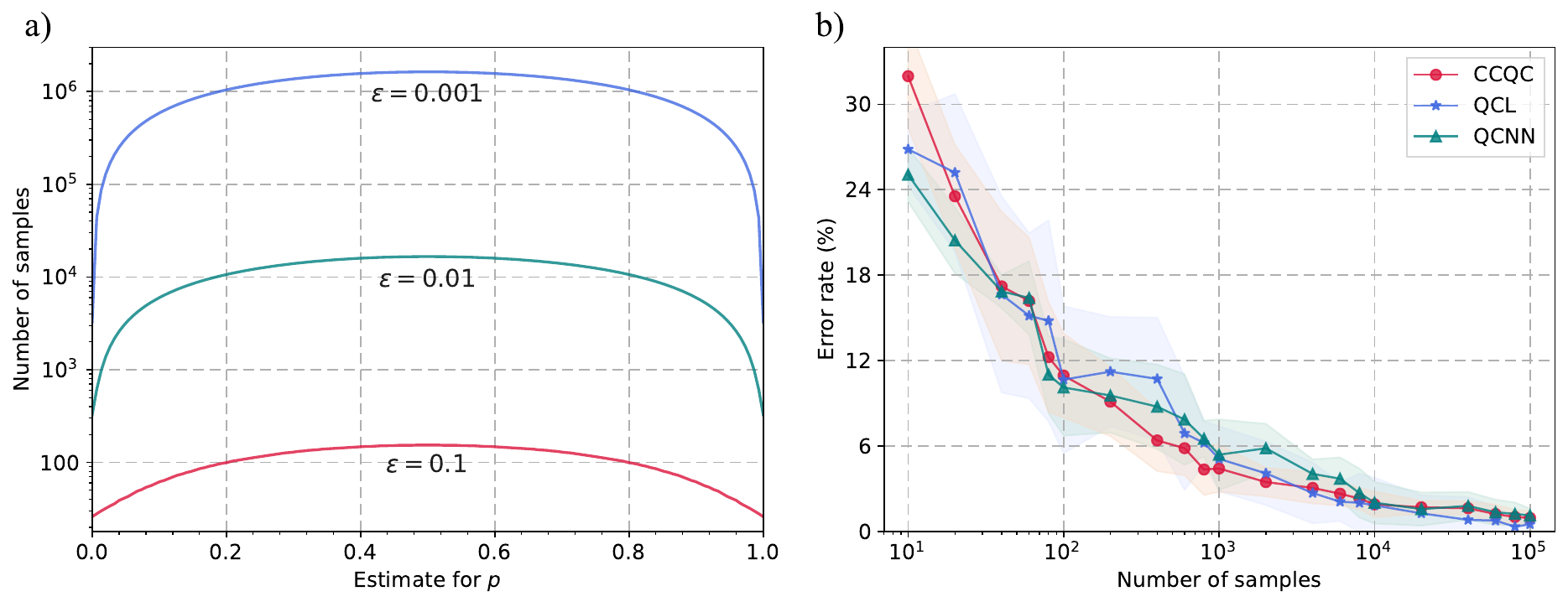}
  \caption{Effect of sampling cost on QuanTest. (a) Relationship between the sampling cost and the mean value $\bar{p}$ for different errors $\epsilon$. (b) Error rate of seed sample predictions by the QNN system under different sampling costs.}
  \Description{.}
  \label{fig:10}
\end{figure*}

The most straightforward and intuitive method to obtain an estimate $\hat{p}$ for $p$ in numerical simulations is maximum likelihood estimation, with the result $\bar{p}$ being the average of all sampled results. 
However, in real scenarios, what concerns us more is how many times $N$ we need to sample from the single qubit measurements in order to estimate the confidence interval $[\hat{p}-\epsilon, \hat{p}+\epsilon]$ for $p$ with a confidence level of $1-\alpha$.
A refined estimation approach is the Wilson score interval, and the error $\epsilon$ can be calculated as:
\begin{equation}
	\epsilon =\frac{z_{\alpha/2}}{1+\frac{z^2_{\alpha/2}}{N}}(\frac{\bar{p}(1-\bar{p})}{N}+\frac{z^2_{\alpha/2}}{4N^2})^{\frac12},
\end{equation}
where $z_{\alpha/2}$ is the quantile (right-tail probabilities) of the stand normal distribution. 
With a confidence level of 99\% (i.e., $z_{\alpha/2}\approx2.58$), the relationship between the sampling cost and $\bar{p}$ can be obtained as shown in Figure \ref{fig:10} (a).
It can be seen that the sampling cost reaches its maximum value when $p=0.5$. 

The test samples generated by QuanTest can detect erroneous behavior in the QNN system. 
When the adversarial examples cause changes in the predicted labels of the tested QNN model, QuanTest will cease to apply perturbations to the samples and output the samples that have passed the quality evaluation as test samples (Algorithm \ref{algorithm1} lines 15-18). 
At this point, the tested QNN model's probability estimate of predicting $y=1$ for the test samples is close to 0.5.
This means that the sampling cost required by the tested QNN system to predict the test samples generated by QuanTest with high confidence will increase significantly,
especially in the face of the real system with fixed and limited sampling times, the test samples are more likely to trigger the system to make incorrect predictions, allowing QuanTest to test the robustness of QNN systems at multiple stages.
the generated samples are more likely to cause the system to misestimate the prediction results.

To verify this, we first mark as seed samples 500 intermediate samples that were iterated 5 times in QuanTest but still failed to flip the model predictions, and then explore the influence of these seed samples on the prediction accuracy of QNN systems under different sampling times.
We chose to evaluate the ternary classification task of MNIST with a low test sample generation rate to make it easier for us to obtain sufficient seed samples.
Note that these seed samples cannot flip the prediction of the model in the case of ideal expectations (i.e., the number of samples tends to be $\infty$). 
Figure \ref{fig:10} (b) shows the error rate of three QNN models as the number of samples varies from $10^1$ to $10^5$, we repeated the classification for each model 10 times to mitigate randomness, where solid lines represent the average results and shadows represent the standard deviation. 
The results confirm our modeling-based inference that when the number of samples is small, the QNN system is more likely to exhibit erroneous behavior, which may also lead to QuanTest generating more test samples than expected.

\begin{figure*}[t]
  \centering
  \includegraphics[width=\linewidth]{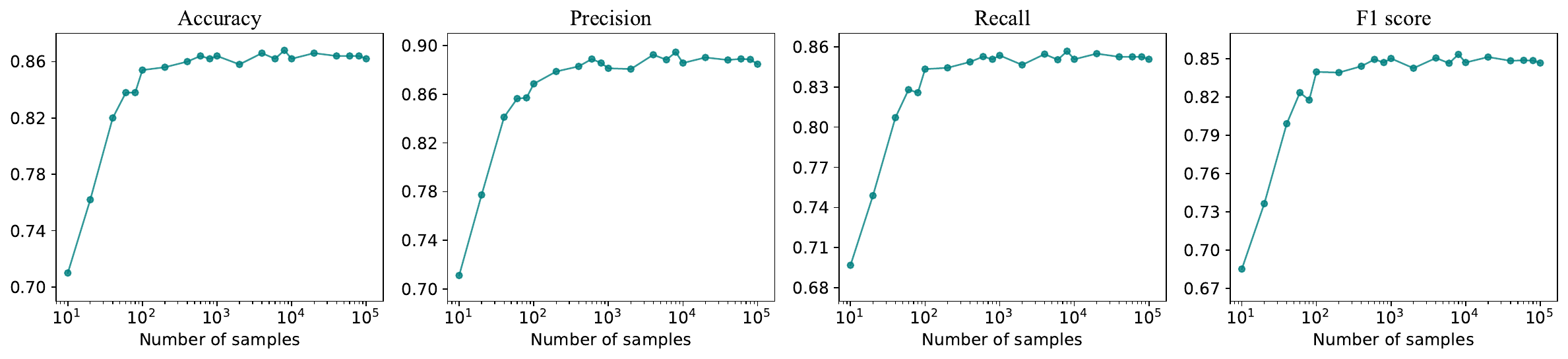}
  \caption{Effect of QuanTest on the quality indicators of the CCQC model under different sampling costs.}
  \Description{.}
  \label{fig:11}
\end{figure*}

To further assess the false negatives and false positives introduced by reducing the number of samples, 
we additionally explored the differences in the effect of QuanTest on the quality indicators of the QNN system under different levels of sampling costs. 
The classification task is set to the ternary classification on MNIST, and the number of iterations for QuanTest is set to 5.
In Figure \ref{fig:11}, we plot the accuracy, precision, recall, and F1 score of the CCQC model as a function of sampling cost.
It is clear from this figure that the quality indicators of the QNN model are much more susceptible to the influence of QuanTest when the number of samples is small. 
As the sampling cost increases, the quality indicators rapidly rise in a similar trend, and when the number of samples exceeds $10^4$, their values stabilize close to the ideal expectations.
This provides us with a reference for balancing physical resources and accurate testing. Considering the results in Figure \ref{fig:10}, setting the number of samples to around $10^4$ during testing might be a relatively balanced choice in this case.

\begin{center}
\begin{tcolorbox}[colback=gray!10,
                  colframe=black,
                  width=\linewidth,
                  arc=1mm, auto outer arc,
                  boxrule=0.5pt,
                 ]
{\bf Answer to RQ4:} The test samples generated by QuanTest can increase the sampling cost of the tested QNN system. When the number of samples is small, QuanTest may generate more test samples than expected, and the quality indicators of the QNN model are more susceptible to the influence of QuanTest.
\end{tcolorbox}
\end{center}

\subsection{Comparison with Black-box Baselines}

The quantum-adapted adversarial attack strategies used by QuanTest have already been empirically evaluated in the classical world independent of entanglement guidance. 
To provide more empirical evidence of effectiveness, we compared QuanTest with black-box adversarial attacks on the binary classification task of Fashion-MNIST, following the same experimental setup as in RQ1.
We select the state-of-the-art black-box adversarial attack method SimBA as the black-box baseline.
Since it is difficult to achieve satisfactory results on the QNN system by directly using the default parameters, we use $\epsilon=0.1$ as the fixed step size of SimBA to compare more fairly with QuanTest.
It should be noted that there is a significant difference between quantum states and classical images.
Modifying quantum states as finely as manipulating classical images in pixel space is difficult. 
Therefore, SimBA here still processes classical data, without quantum-based adjustments.

Table~\ref{table:5.2} shows the comparison between QuanTest and the black-box baseline.
It can be seen that although the black box baseline performs better than the random noise, there is still a significant gap compared to QuanTest. Unlike QuanTest, the black-box baseline finds it more difficult to generate test samples on the Fashion dataset, possibly because black-box attacks directly process classical images, leading to performance differences between different datasets.
It is worth emphasizing that the QNN model can classify classical and quantum data \cite{QML}. 
Adding perturbations to classical data is equivalent to adding perturbations to the corresponding quantum states \cite{QAML1}, so QuanTest only needs to consider the usage scenarios for quantum data.
End-to-end black-box adversarial attacks aimed at classical data cannot be directly applied to quantum data. Therefore, the usage scenarios of QuanTest and end-to-end black-box attacks are different.

\begin{table}
\belowrulesep=0.15pt
\aboverulesep=0.15pt
  \caption{Comparison of QuanTest and black-box baseline on test sample generation rate and average trace distance.}
  \label{table:5.2}
\resizebox{0.67\columnwidth}{!}{
\renewcommand{\arraystretch}{1.25}
\begin{tabular}{@{}cccccc@{}}

\toprule
\multirow{2}{*}{Data} & \multirow{2}{*}{Model}  & \multicolumn{2}{|c|}{Gen\_Rate(\%)} & \multicolumn{2}{c}{ATD} \\\cmidrule(lr){3-4}  \cmidrule(lr){5-6}
                                     &                           &  \multicolumn{1}{|c}{SimBA}  & \multicolumn{1}{c|} {\makecell{QuanTest\\(DLFuzz-based)}}  & SimBA    & {\makecell{QuanTest\\(DLFuzz-based)}}\\
\midrule
\multirow{3}{*}{Fashion[1,4]}          & {CCQC}            & \multicolumn{1}{|c}{7.43}    & \multicolumn{1}{c|}{\bf87.79}    & \bf0.3136    & {0.3428}  \\
                                     &{QCL}                   & \multicolumn{1}{|c}{7.57}    & \multicolumn{1}{c|}{\bf96.48}   & \bf0.3224 & 0.3327  \\
                                     & {QCNN}             & \multicolumn{1}{|c}{13.71}    & \multicolumn{1}{c|}{\bf92.73}   & 0.3370    & \bf0.2947   \\
\midrule
\multirow{3}{*}{MNIST[0,1,2]}          & {CCQC}            & \multicolumn{1}{|c}{16.23}    & \multicolumn{1}{c|}{\bf70.95}    & 0.3261    & \bf0.2852  \\
                                     &{QCL}                   & \multicolumn{1}{|c}{19.11}    & \multicolumn{1}{c|}{\bf71.52}   & 0.3313 & \bf0.2467  \\
                                     & {QCNN}             & \multicolumn{1}{|c}{24.51}    & \multicolumn{1}{c|}{\bf79.24}   & 0.3294    & \bf0.1984   \\
\midrule   
\multirow{3}{*}{MNIST[3,6]}          & {CCQC}            & \multicolumn{1}{|c}{13.54}    & \multicolumn{1}{c|}{\bf88.44}    & 0.3588    & \bf0.3117  \\
                                     &{QCL}                   & \multicolumn{1}{|c}{21.86}    & \multicolumn{1}{c|}{\bf91.70}   & 0.3521 & \bf0.2811  \\
                                     & {QCNN}             & \multicolumn{1}{|c}{20.68}    & \multicolumn{1}{c|}{\bf96.87}   & 0.3462    & \bf0.2881   \\
\bottomrule
\end{tabular}
}
\end{table}

\begin{center}
\begin{tcolorbox}[colback=gray!10,
                  colframe=black,
                  width=\linewidth,
                  arc=1mm, auto outer arc,
                  boxrule=0.5pt,
                 ]
{\bf Answer to RQ5:} QuanTest has a better performance than the black box baseline. Note that QuanTest and end-to-end black box attacks have different usage scenarios.
\end{tcolorbox}
\end{center}

\subsection{Selection of Random Baseline}

As existing work on testing QNN systems is still blank, we selected random coherent noise as the baseline to evaluate QuanTest in RQ1.
Considering that random noise of smaller strength per iteration may have a higher test sample generation rate, we explore the generation of test samples using random coherent noise at different $\sigma$ settings on the Fashion dataset, following the same experimental settings in RQ1. The results are shown in Table \ref{table:5.1}.

We find that smaller values of $\sigma$ indeed lead to a higher generation rate of random noise, but this still has a significant gap compared to QuanTest's performance. However, it is imperative to note that the decrease in strength means an increase in the number of iterations, which will result in greater time overhead. In fact, within the same time cost as QuanTest, more test samples are generated by random noise with $\sigma=0.02$ than other strength settings.
Additionally, the time required for random noise with $\sigma=0.02$ to reach the same similarity limit as QuanTest is also closer. So it is a fairer setting to select the random noise of $\sigma=0.02$ as the baseline.

\begin{table}
\belowrulesep=0.15pt
\aboverulesep=0.15pt
  \caption{Comparison of random baselines with different noise strengths on test sample generation rate and average trace distance.}
  \label{table:5.1}
\resizebox{0.67\columnwidth}{!}{
\renewcommand{\arraystretch}{1.25}
\begin{tabular}{@{}ccccccc@{}}

\toprule
 \multirow{2}{*}{Model}  & \multicolumn{3}{|c|}{Gen\_Rate(\%)} & \multicolumn{3}{c}{ATD} \\\cmidrule(lr){2-4}  \cmidrule(lr){5-7}
                        &  \multicolumn{1}{|c}{$\sigma=0.02$}  & \multicolumn{1}{c} {$\sigma=0.01$}  &\multicolumn{1}{c|} {$\sigma=0.005$}  & $\sigma=0.02$    & {$\sigma=0.01$} &\multicolumn{1}{c}{$\sigma=0.005$} \\
\midrule
      {CCQC}  & \multicolumn{1}{|c}{0.62}    & \multicolumn{1}{c}{0.72} & \multicolumn{1}{c|}{\bf1.52}   & \bf0.2143    & {0.2517}  & 0.2727 \\
      {QCL}     & \multicolumn{1}{|c}{0.83}    & \multicolumn{1}{c}{1.18} & \multicolumn{1}{c|}{\bf1.68}   & 0.3138    & 0.2733   & \bf0.2503 \\
      {QCNN}  & \multicolumn{1}{|c}{2.49}    & \multicolumn{1}{c}{2.13}  & \multicolumn{1}{c|}{2.32}     &  \bf0.2669      & \bf0.2696  & 0.2713 \\

\bottomrule
\end{tabular}
}
\end{table}

\subsection{Scalability}

The adversarial attack module used in QuanTest is configurable, which means that QuanTest can scale various adversarial attack strategies to generate diverse test samples without the need for quantum adaptation for each strategy. 
However, this scalability may be limited in end-to-end black-box attack strategies, as effective black-box attack methods specifically targeting quantum states have not yet been proposed. 
Additionally, based on our entire evaluation, QuanTest also has the ability to scale to high-dimensional quantum systems (i.e., larger networks and datasets) and a greater number of classes. 
Therefore, QuanTest's testing targets are not limited to QNN systems designed for current NISQ. In future work, QuanTest will attempt to extend testing tasks to a broader range of architectures. 
It is worth noting our proposed QEA metric, as it relies solely on entanglement measure, theoretically QEA can be extended to other structural coverage test objectives for QNN systems.

\subsection {Threats to Validity}
Quantum data encoding is a major threat to validity.
Current classical computers can only simulate a limited number of qubits due to the enormous computational cost incurred by numerical simulations of high-dimensional quantum systems \cite{QML}.
Classical data is typically downsampled and then encoded into quantum states for training QNNs \cite{QCNNimg}.
We alleviate this threat by using two popular datasets with lower data dimensions Fashion and MNIST, for evaluation.

Another threat could be the selection of QNN models. 
The construction of QNNs is still under exploration,
and there exist various QNN variants based on different underlying architectures \cite{QONN, CVQNN}.
To ensure the generalizability of QuanTest, we employ three different architectures of QNN models to evaluate QuanTest's performance.
We believe that QuanTest can be applied to the majority of PQC-based QNNs, and the testing of other QNN systems with different architectures is worth exploring in future work.

Finally, variations in quantum computer implementation technologies also pose a potential threat.
To guarantee the applicability of QuanTest in the current NISQ era, all methods we use are based on demonstrations on real devices. 
Specifically, fidelity, as a similarity evaluation metric, can be obtained by performing the SWAP test.
For the core part of QuanTest, Ren et al. \cite{QAML2} demonstrated the experimental generation of quantum adversarial examples using an array of ten programmable superconducting transmon qubits. 
Similarly, Haug et al. \cite{MWreal} demonstrated the Bell measurement protocol for MW entanglement measurement on the IonQ quantum computer.

\section{Related Works}

\subsection{Deep Learning Testing}
Traditional black-box testing paradigms, such as manually labeled data and simulations, may not be able to discover various extreme cases that can lead to erroneous behaviors in machine learning systems. 
Pei et al. \cite{Deepxplore} innovatively proposed a white-box differential testing method, 
introducing neuron coverage as the first testing criterion to measure the extent to which the internal logic of DNN is being tested. 
Subsequently, several related testing works based on neuron coverage emerged. 
Deeptest \cite{Deeptest} automatically generates test samples by applying various real image transformations (e.g., rotation and blurring) to maximize neuron coverage, and detect numerous potential defect behaviors in DNN-driven autonomous driving systems.
DLFuzz \cite{DLFuzz} employs fuzzing into DL testing, and keeps minute mutation to the original samples, thereby maximizing the neuron coverage of mutated samples and the prediction differences from the quantum original samples.

Several new testing criteria have also been proposed since then. Deepgauge \cite{Deepgauge} proposed a set of more fine-grained neuron coverage criteria. 
DeepCT \cite{DeepCT} introduced a set of coverage metrics for combinatorial testing based on neuron layers. 
However, the capacity of neuron coverage and its variants in test sample generation has also been questioned in some studies \cite{QuestionNC_1, QuestionNC_2}. 
In addition to coverage-guided testing, guiding testing using surprise adequacy \cite{Guidingloss} or robustness \cite{Robot} have also been proposed, as well as some mutation testing \cite{DeepMutation, Mutation2, DeepCrime} for DL systems.
The unique features of RNN models make the work of RNN testing \cite{Deepstellar, Deepstate, RNNTest} distinctive. 
Deepstellar \cite{Deepstellar} abstracted an RNN into discrete-time Markov chain models and designed five coverage criteria based on this abstract model's state transitions. 
RNN-Test \cite{RNNTest} designed two coverage metrics for RNNs from the perspectives of state coverage.
However, the work of QNN systems testing has garnered limited attention thus far. 
Due to its fundamentally different computational paradigm from classical computing,
existing testing methods and criteria may not be suitable for it.

\subsection{Adversarial Machine Learning}
The vulnerability of machine learning models has garnered considerable attention \cite{Advattention1, Advattention2}.
Attackers can deceive DNNs by applying imperceptible perturbations to input images.
At present, there are plenty of works on conducting adversarial attacks on DL systems to generate adversarial examples \cite{FGSM, BIM, CW, PGD, JSMA, DeepFool}.
adversarial examples existing in various input data such as images and text pose a broad threat to learning models \cite{am10, attackIMG, attackTEXT}.

In recent years, vulnerabilities to adversarial examples have also been found in quantum machine learning models \cite{QAML1, QAML3}.
Quantum adversarial learning has been experimentally demonstrated on programmable superconducting qubits \cite{QAML2}.
In contrast to these works, QuanTest systematically tests QNN systems by entanglement guidance.
Moreover, QuanTest is exceptionally scalable and applicable to various PQC-based QNN systems and quantum state encoding strategies.
Modularity allows QuanTest to directly achieve most quantum-adapted adversarial attack methods without the need for additional adjustments and adaptations, thus generating various quantum adversarial examples with high efficiency.

\subsection{Quantum Software Testing}
With the rapid advancements in scale and reliability of quantum computing \cite{QCdevelopment1, QCdevelopment2, QCdevelopment3}, 
the demand for quantum software development and testing is growing.
Companies like Google\footnote{https://quantumai.google/} and IBM\footnote{https://www.ibm.com/quantum}, while building their own quantum computers, 
have also introduced a range of quantum software development tools and quantum cloud platforms, such as Qiskit \cite{Qiskit}, PennyLane \cite{PennyLane}, Cirq \cite{Cirq}, and TensorFlow Quantum \cite{TensorflowQuantum}. 
These tools aim to assist engineers and researchers in developing quantum programs that can run on quantum computers.

At present, quantum software testing is largely still in its infancy.
Due to the specialized characteristics of quantum computing, such as entanglement and superposition \cite{QCQI}, quantum software testing faces challenges in various aspects, including state readout.
Automated and systematic testing methods that can effectively scale for complex quantum programs on quantum computers are still missing due to factors like quantum noise, the scale of qubits, and underlying implementations.
Most quantum software testing methods attempt to apply some of the testing techniques already existing in the classical world, 
such as differential testing \cite{QSEdiff}, fuzz testing \cite{QSE1}, mutation testing \cite{QSE2, QSEmut1, QSEmut2}, and coverage criteria \cite{QSEcov1, QSEcov2}, 
while making necessary adaptations to accommodate the inherent properties of quantum systems.
Nevertheless, QNNs have a different programming paradigm and decision logic representation compared to traditional quantum software.
Just as traditional software testing techniques could not be directly applied to classic DL systems,
existing quantum software testing methods are equally inadequate for QNN systems.

\section{Conclusion}
In this paper, we propose QuanTest, an adversarial testing framework for QNN systems. 
QuanTest includes a set of similarity metrics and an entanglement adequacy criterion that can effectively detect erroneous behaviors in QNN systems under entanglement guidance, 
generating high-quality adversarial examples. 
Our experimental results demonstrate the effectiveness of QuanTest, where QuanTest is capable of generating several hundred times more test samples than random noise under equivalent perturbation size constraints. 
The average fidelity measure of these test samples consistently exceeds 95\%.

\begin{acks}

We thank Jia Liao and Yuhu Lu for assistance with the writing and revision process. 
\end{acks}


\bibliographystyle{ACM-Reference-Format}
\bibliography{QuanTest}


\begin{thebibliography}{82}


\ifx \showCODEN    \undefined \def \showCODEN     #1{\unskip}     \fi
\ifx \showDOI      \undefined \def \showDOI       #1{#1}\fi
\ifx \showISBNx    \undefined \def \showISBNx     #1{\unskip}     \fi
\ifx \showISBNxiii \undefined \def \showISBNxiii  #1{\unskip}     \fi
\ifx \showISSN     \undefined \def \showISSN      #1{\unskip}     \fi
\ifx \showLCCN     \undefined \def \showLCCN      #1{\unskip}     \fi
\ifx \shownote     \undefined \def \shownote      #1{#1}          \fi
\ifx \showarticletitle \undefined \def \showarticletitle #1{#1}   \fi
\ifx \showURL      \undefined \def \showURL       {\relax}        \fi
\providecommand\bibfield[2]{#2}
\providecommand\bibinfo[2]{#2}
\providecommand\natexlab[1]{#1}
\providecommand\showeprint[2][]{arXiv:#2}

\bibitem[QCd(2023)]%
        {QCdevelopment1}
 \bibinfo{year}{2023}\natexlab{}.
\newblock \showarticletitle{Suppressing quantum errors by scaling a surface code logical qubit}.
\newblock \bibinfo{journal}{\emph{Nature}} \bibinfo{volume}{614}, \bibinfo{number}{7949} (\bibinfo{year}{2023}), \bibinfo{pages}{676--681}.
\newblock


\bibitem[Ali et~al\mbox{.}(2021)]%
        {QSEcov2}
\bibfield{author}{\bibinfo{person}{Shaukat Ali}, \bibinfo{person}{Paolo Arcaini}, \bibinfo{person}{Xinyi Wang}, {and} \bibinfo{person}{Tao Yue}.} \bibinfo{year}{2021}\natexlab{}.
\newblock \showarticletitle{Assessing the Effectiveness of Input and Output Coverage Criteria for Testing Quantum Programs}. In \bibinfo{booktitle}{\emph{2021 14th IEEE Conference on Software Testing, Verification and Validation (ICST)}}. \bibinfo{publisher}{IEEE}, \bibinfo{pages}{13--23}.
\newblock


\bibitem[Ali and Yue(2023)]%
        {QSE0}
\bibfield{author}{\bibinfo{person}{Shaukat Ali} {and} \bibinfo{person}{Tao Yue}.} \bibinfo{year}{2023}\natexlab{}.
\newblock \showarticletitle{Quantum Software Testing: A Brief Introduction}. In \bibinfo{booktitle}{\emph{2023 IEEE/ACM 45th International Conference on Software Engineering: Companion Proceedings (ICSE-Companion)}}. \bibinfo{publisher}{IEEE}, \bibinfo{pages}{332--333}.
\newblock


\bibitem[Anil et~al\mbox{.}(2024)]%
        {QAML13}
\bibfield{author}{\bibinfo{person}{Gautham Anil}, \bibinfo{person}{Vishnu Vinod}, {and} \bibinfo{person}{Apurva Narayan}.} \bibinfo{year}{2024}\natexlab{}.
\newblock \bibinfo{title}{Generating Universal Adversarial Perturbations for Quantum Classifiers}.
\newblock , \bibinfo{numpages}{10891-10899}~pages.
\newblock
\urldef\tempurl%
\url{https://doi.org/10.1609/aaai.v38i10.28963}
\showDOI{\tempurl}


\bibitem[Arrazola et~al\mbox{.}(2021)]%
        {QCdevelopment3}
\bibfield{author}{\bibinfo{person}{Juan~M Arrazola}, \bibinfo{person}{Ville Bergholm}, \bibinfo{person}{Kamil Br{\'a}dler}, \bibinfo{person}{Thomas~R Bromley}, \bibinfo{person}{Matt~J Collins}, \bibinfo{person}{Ish Dhand}, \bibinfo{person}{Alberto Fumagalli}, \bibinfo{person}{Thomas Gerrits}, \bibinfo{person}{Andrey Goussev}, \bibinfo{person}{Lukas~G Helt}, {et~al\mbox{.}}} \bibinfo{year}{2021}\natexlab{}.
\newblock \showarticletitle{Quantum circuits with many photons on a programmable nanophotonic chip}.
\newblock \bibinfo{journal}{\emph{Nature}} \bibinfo{volume}{591}, \bibinfo{number}{7848} (\bibinfo{year}{2021}), \bibinfo{pages}{54--60}.
\newblock


\bibitem[Ball(2021)]%
        {QCdevelopment2}
\bibfield{author}{\bibinfo{person}{Philip Ball}.} \bibinfo{year}{2021}\natexlab{}.
\newblock \showarticletitle{First 100-QUBIT quantum computer enters crowded race}.
\newblock \bibinfo{journal}{\emph{Nature}}  \bibinfo{volume}{599} (\bibinfo{year}{2021}), \bibinfo{pages}{542}.
\newblock


\bibitem[Bausch(2020)]%
        {QRNN}
\bibfield{author}{\bibinfo{person}{Johannes Bausch}.} \bibinfo{year}{2020}\natexlab{}.
\newblock \showarticletitle{Recurrent quantum neural networks}.
\newblock \bibinfo{journal}{\emph{Advances in neural information processing systems}}  \bibinfo{volume}{33} (\bibinfo{year}{2020}), \bibinfo{pages}{1368--1379}.
\newblock
\urldef\tempurl%
\url{https://doi.org/10.48550/arXiv.2006.14619}
\showDOI{\tempurl}


\bibitem[Benedetti et~al\mbox{.}(2019)]%
        {PQC}
\bibfield{author}{\bibinfo{person}{Marcello Benedetti}, \bibinfo{person}{Erika Lloyd}, \bibinfo{person}{Stefan Sack}, {and} \bibinfo{person}{Mattia Fiorentini}.} \bibinfo{year}{2019}\natexlab{}.
\newblock \showarticletitle{Parameterized quantum circuits as machine learning models}.
\newblock \bibinfo{journal}{\emph{Quantum Science and Technology}} \bibinfo{volume}{4}, \bibinfo{number}{4} (\bibinfo{date}{nov} \bibinfo{year}{2019}), \bibinfo{pages}{043001}.
\newblock
\urldef\tempurl%
\url{https://doi.org/10.1088/2058-9565/ab4eb5}
\showDOI{\tempurl}


\bibitem[Bergholm et~al\mbox{.}(2018)]%
        {PennyLane}
\bibfield{author}{\bibinfo{person}{Ville Bergholm}, \bibinfo{person}{Josh~A. Izaac}, \bibinfo{person}{Maria Schuld}, \bibinfo{person}{Christian Gogolin}, {and} \bibinfo{person}{Nathan Killoran}.} \bibinfo{year}{2018}\natexlab{}.
\newblock \showarticletitle{PennyLane: Automatic differentiation of hybrid quantum-classical computations}.
\newblock \bibinfo{journal}{\emph{CoRR}}  \bibinfo{volume}{abs/1811.04968} (\bibinfo{year}{2018}).
\newblock
\showeprint{1811.04968}
\urldef\tempurl%
\url{http://arxiv.org/abs/1811.04968}
\showURL{%
\tempurl}


\bibitem[Biamonte et~al\mbox{.}(2017)]%
        {QML}
\bibfield{author}{\bibinfo{person}{Jacob Biamonte}, \bibinfo{person}{Peter Wittek}, \bibinfo{person}{Nicola Pancotti}, \bibinfo{person}{Patrick Rebentrost}, \bibinfo{person}{Nathan Wiebe}, {and} \bibinfo{person}{Seth Lloyd}.} \bibinfo{year}{2017}\natexlab{}.
\newblock \showarticletitle{Quantum machine learning}.
\newblock \bibinfo{journal}{\emph{Nature}} \bibinfo{volume}{549}, \bibinfo{number}{7671} (\bibinfo{date}{01 Sep} \bibinfo{year}{2017}), \bibinfo{pages}{195--202}.
\newblock
\showISSN{1476-4687}
\urldef\tempurl%
\url{https://doi.org/10.1038/nature23474}
\showDOI{\tempurl}


\bibitem[Biggio and Roli(2018)]%
        {am10}
\bibfield{author}{\bibinfo{person}{Battista Biggio} {and} \bibinfo{person}{Fabio Roli}.} \bibinfo{year}{2018}\natexlab{}.
\newblock \showarticletitle{Wild Patterns: Ten Years After the Rise of Adversarial Machine Learning}. In \bibinfo{booktitle}{\emph{Proceedings of the 2018 ACM SIGSAC Conference on Computer and Communications Security}} (Toronto, Canada) \emph{(\bibinfo{series}{CCS '18})}. \bibinfo{publisher}{Association for Computing Machinery}, \bibinfo{address}{New York, NY, USA}, \bibinfo{pages}{2154–2156}.
\newblock
\urldef\tempurl%
\url{https://doi.org/10.1145/3243734.3264418}
\showDOI{\tempurl}


\bibitem[Broughton et~al\mbox{.}(2020)]%
        {TensorflowQuantum}
\bibfield{author}{\bibinfo{person}{Michael Broughton}, \bibinfo{person}{Guillaume Verdon}, \bibinfo{person}{Trevor McCourt}, \bibinfo{person}{Antonio~J Martinez}, \bibinfo{person}{Jae~Hyeon Yoo}, \bibinfo{person}{Sergei~V Isakov}, \bibinfo{person}{Philip Massey}, \bibinfo{person}{Ramin Halavati}, \bibinfo{person}{Murphy~Yuezhen Niu}, \bibinfo{person}{Alexander Zlokapa}, {et~al\mbox{.}}} \bibinfo{year}{2020}\natexlab{}.
\newblock \showarticletitle{Tensorflow quantum: A software framework for quantum machine learning}.
\newblock \bibinfo{journal}{\emph{arXiv preprint arXiv:2003.02989}} (\bibinfo{year}{2020}).
\newblock


\bibitem[Carlini and Wagner(2017)]%
        {CW}
\bibfield{author}{\bibinfo{person}{Nicholas Carlini} {and} \bibinfo{person}{David Wagner}.} \bibinfo{year}{2017}\natexlab{}.
\newblock \showarticletitle{Towards evaluating the robustness of neural networks}. In \bibinfo{booktitle}{\emph{2017 ieee symposium on security and privacy (sp)}}. \bibinfo{publisher}{IEEE}, \bibinfo{pages}{39--57}.
\newblock


\bibitem[Cerezo et~al\mbox{.}(2021)]%
        {VQA}
\bibfield{author}{\bibinfo{person}{Marco Cerezo}, \bibinfo{person}{Andrew Arrasmith}, \bibinfo{person}{Ryan Babbush}, \bibinfo{person}{Simon~C Benjamin}, \bibinfo{person}{Suguru Endo}, \bibinfo{person}{Keisuke Fujii}, \bibinfo{person}{Jarrod~R McClean}, \bibinfo{person}{Kosuke Mitarai}, \bibinfo{person}{Xiao Yuan}, \bibinfo{person}{Lukasz Cincio}, {et~al\mbox{.}}} \bibinfo{year}{2021}\natexlab{}.
\newblock \showarticletitle{Variational quantum algorithms}.
\newblock \bibinfo{journal}{\emph{Nature Reviews Physics}} \bibinfo{volume}{3}, \bibinfo{number}{9} (\bibinfo{date}{aug} \bibinfo{year}{2021}), \bibinfo{pages}{625--644}.
\newblock
\urldef\tempurl%
\url{https://doi.org/10.1038/s42254-021-00348-9}
\showDOI{\tempurl}


\bibitem[Cong et~al\mbox{.}(2019)]%
        {cong2019quantum}
\bibfield{author}{\bibinfo{person}{Iris Cong}, \bibinfo{person}{Soonwon Choi}, {and} \bibinfo{person}{Mikhail~D. Lukin}.} \bibinfo{year}{2019}\natexlab{}.
\newblock \showarticletitle{Quantum convolutional neural networks}.
\newblock \bibinfo{journal}{\emph{Nature Physics}} \bibinfo{volume}{15}, \bibinfo{number}{12} (\bibinfo{date}{01 Dec} \bibinfo{year}{2019}), \bibinfo{pages}{1273--1278}.
\newblock
\showISSN{1745-2481}
\urldef\tempurl%
\url{https://doi.org/10.1038/s41567-019-0648-8}
\showDOI{\tempurl}


\bibitem[Cross(2018)]%
        {Qiskit}
\bibfield{author}{\bibinfo{person}{Andrew Cross}.} \bibinfo{year}{2018}\natexlab{}.
\newblock \showarticletitle{The IBM Q experience and QISKit open-source quantum computing software}. In \bibinfo{booktitle}{\emph{APS March meeting abstracts}}, Vol.~\bibinfo{volume}{2018}. \bibinfo{pages}{L58--003}.
\newblock


\bibitem[Developers(2023)]%
        {Cirq}
\bibfield{author}{\bibinfo{person}{Cirq Developers}.} \bibinfo{year}{2023}\natexlab{}.
\newblock \bibinfo{booktitle}{\emph{Cirq}}.
\newblock
\urldef\tempurl%
\url{https://doi.org/10.5281/zenodo.8161252}
\showDOI{\tempurl}


\bibitem[Du et~al\mbox{.}(2019)]%
        {Deepstellar}
\bibfield{author}{\bibinfo{person}{Xiaoning Du}, \bibinfo{person}{Xiaofei Xie}, \bibinfo{person}{Yi Li}, \bibinfo{person}{Lei Ma}, \bibinfo{person}{Yang Liu}, {and} \bibinfo{person}{Jianjun Zhao}.} \bibinfo{year}{2019}\natexlab{}.
\newblock \showarticletitle{Deepstellar: Model-based quantitative analysis of stateful deep learning systems}. In \bibinfo{booktitle}{\emph{Proceedings of the 2019 27th ACM Joint Meeting on European Software Engineering Conference and Symposium on the Foundations of Software Engineering}}. \bibinfo{pages}{477--487}.
\newblock


\bibitem[Fortunato et~al\mbox{.}(2022a)]%
        {QSEmut2}
\bibfield{author}{\bibinfo{person}{Daniel Fortunato}, \bibinfo{person}{Jos\'{e} Campos}, {and} \bibinfo{person}{Rui Abreu}.} \bibinfo{year}{2022}\natexlab{a}.
\newblock \showarticletitle{Mutation testing of quantum programs written in QISKit}. In \bibinfo{booktitle}{\emph{Proceedings of the ACM/IEEE 44th International Conference on Software Engineering: Companion Proceedings}} (Pittsburgh, Pennsylvania) \emph{(\bibinfo{series}{ICSE '22})}. \bibinfo{publisher}{Association for Computing Machinery}, \bibinfo{address}{New York, NY, USA}, \bibinfo{pages}{358–359}.
\newblock
\urldef\tempurl%
\url{https://doi.org/10.1145/3510454.3528649}
\showDOI{\tempurl}


\bibitem[Fortunato et~al\mbox{.}(2022b)]%
        {QSEmut1}
\bibfield{author}{\bibinfo{person}{Daniel Fortunato}, \bibinfo{person}{Jos\'{e} Campos}, {and} \bibinfo{person}{Rui Abreu}.} \bibinfo{year}{2022}\natexlab{b}.
\newblock \showarticletitle{QMutPy: a mutation testing tool for Quantum algorithms and applications in Qiskit}. In \bibinfo{booktitle}{\emph{Proceedings of the 31st ACM SIGSOFT International Symposium on Software Testing and Analysis}} \emph{(\bibinfo{series}{ISSTA 2022})}. \bibinfo{publisher}{Association for Computing Machinery}, \bibinfo{address}{New York, NY, USA}, \bibinfo{pages}{797–800}.
\newblock
\urldef\tempurl%
\url{https://doi.org/10.1145/3533767.3543296}
\showDOI{\tempurl}


\bibitem[Gong and Deng(2022)]%
        {QAML3}
\bibfield{author}{\bibinfo{person}{Weiyuan Gong} {and} \bibinfo{person}{Dong-Ling Deng}.} \bibinfo{year}{2022}\natexlab{}.
\newblock \showarticletitle{Universal adversarial examples and perturbations for quantum classifiers}.
\newblock \bibinfo{journal}{\emph{National Science Review}} \bibinfo{volume}{9}, \bibinfo{number}{6} (\bibinfo{year}{2022}), \bibinfo{pages}{nwab130}.
\newblock


\bibitem[Goodfellow et~al\mbox{.}(2015)]%
        {FGSM}
\bibfield{author}{\bibinfo{person}{Ian Goodfellow}, \bibinfo{person}{Jonathon Shlens}, {and} \bibinfo{person}{Christian Szegedy}.} \bibinfo{year}{2015}\natexlab{}.
\newblock \showarticletitle{Explaining and Harnessing Adversarial Examples}. In \bibinfo{booktitle}{\emph{International Conference on Learning Representations}}. \bibinfo{publisher}{IEEE}, \bibinfo{address}{New York, NY, USA}.
\newblock
\urldef\tempurl%
\url{https://doi.org/10.48550/arXiv.1412.6572}
\showDOI{\tempurl}


\bibitem[Guo et~al\mbox{.}(2018)]%
        {DLFuzz}
\bibfield{author}{\bibinfo{person}{Jianmin Guo}, \bibinfo{person}{Yu Jiang}, \bibinfo{person}{Yue Zhao}, \bibinfo{person}{Quan Chen}, {and} \bibinfo{person}{Jiaguang Sun}.} \bibinfo{year}{2018}\natexlab{}.
\newblock \showarticletitle{DLFuzz: Differential Fuzzing Testing of Deep Learning Systems}. In \bibinfo{booktitle}{\emph{Proceedings of the 2018 26th ACM Joint Meeting on European Software Engineering Conference and Symposium on the Foundations of Software Engineering}} (Lake Buena Vista, FL, USA). \bibinfo{publisher}{Association for Computing Machinery}, \bibinfo{address}{New York, NY, USA}, \bibinfo{pages}{739–743}.
\newblock
\showISBNx{9781450355735}
\urldef\tempurl%
\url{https://doi.org/10.1145/3236024.3264835}
\showDOI{\tempurl}


\bibitem[Guo et~al\mbox{.}(2022)]%
        {RNNTest}
\bibfield{author}{\bibinfo{person}{Jianmin Guo}, \bibinfo{person}{Quan Zhang}, \bibinfo{person}{Yue Zhao}, \bibinfo{person}{Heyuan Shi}, \bibinfo{person}{Yu Jiang}, {and} \bibinfo{person}{Jia{-}Guang Sun}.} \bibinfo{year}{2022}\natexlab{}.
\newblock \showarticletitle{RNN-Test: Towards Adversarial Testing for Recurrent Neural Network Systems}.
\newblock \bibinfo{journal}{\emph{{IEEE} Trans. Software Eng.}} \bibinfo{volume}{48}, \bibinfo{number}{10} (\bibinfo{date}{sep} \bibinfo{year}{2022}), \bibinfo{pages}{4167--4180}.
\newblock


\bibitem[Harel{-}Canada et~al\mbox{.}(2020)]%
        {QuestionNC_1}
\bibfield{author}{\bibinfo{person}{Fabrice Harel{-}Canada}, \bibinfo{person}{Lingxiao Wang}, \bibinfo{person}{Muhammad~Ali Gulzar}, \bibinfo{person}{Quanquan Gu}, {and} \bibinfo{person}{Miryung Kim}.} \bibinfo{year}{2020}\natexlab{}.
\newblock \showarticletitle{Is neuron coverage a meaningful measure for testing deep neural networks?}. In \bibinfo{booktitle}{\emph{Proceedings of the 28th ACM Joint Meeting on European Software Engineering Conference and Symposium on the Foundations of Software Engineering}}. \bibinfo{publisher}{{ACM}}, \bibinfo{pages}{851--862}.
\newblock


\bibitem[Harrow et~al\mbox{.}(2009)]%
        {HHL}
\bibfield{author}{\bibinfo{person}{Aram~W Harrow}, \bibinfo{person}{Avinatan Hassidim}, {and} \bibinfo{person}{Seth Lloyd}.} \bibinfo{year}{2009}\natexlab{}.
\newblock \showarticletitle{Quantum algorithm for linear systems of equations}.
\newblock \bibinfo{journal}{\emph{Physical review letters}} \bibinfo{volume}{103}, \bibinfo{number}{15} (\bibinfo{year}{2009}), \bibinfo{pages}{150502}.
\newblock


\bibitem[Haug and Kim(2023)]%
        {MWreal}
\bibfield{author}{\bibinfo{person}{Tobias Haug} {and} \bibinfo{person}{M.S. Kim}.} \bibinfo{year}{2023}\natexlab{}.
\newblock \showarticletitle{Scalable Measures of Magic Resource for Quantum Computers}.
\newblock \bibinfo{journal}{\emph{PRX Quantum}}  \bibinfo{volume}{4} (\bibinfo{date}{Jan} \bibinfo{year}{2023}), \bibinfo{pages}{010301}.
\newblock
Issue 1.
\urldef\tempurl%
\url{https://doi.org/10.1103/PRXQuantum.4.010301}
\showDOI{\tempurl}


\bibitem[Havl{\'\i}{\v{c}}ek et~al\mbox{.}(2019)]%
        {havlivcek2019supervised}
\bibfield{author}{\bibinfo{person}{Vojt{\v{e}}ch Havl{\'\i}{\v{c}}ek}, \bibinfo{person}{Antonio~D C{\'o}rcoles}, \bibinfo{person}{Kristan Temme}, \bibinfo{person}{Aram~W Harrow}, \bibinfo{person}{Abhinav Kandala}, \bibinfo{person}{Jerry~M Chow}, {and} \bibinfo{person}{Jay~M Gambetta}.} \bibinfo{year}{2019}\natexlab{}.
\newblock \showarticletitle{Supervised learning with quantum-enhanced feature spaces}.
\newblock \bibinfo{journal}{\emph{Nature}} \bibinfo{volume}{567}, \bibinfo{number}{7747} (\bibinfo{date}{Mar} \bibinfo{year}{2019}), \bibinfo{pages}{209--212}.
\newblock
\showISSN{1476-4687}
\urldef\tempurl%
\url{https://doi.org/10.1038/s41586-019-0980-2}
\showDOI{\tempurl}


\bibitem[Helstrom(1969)]%
        {Trace}
\bibfield{author}{\bibinfo{person}{Carl~W. Helstrom}.} \bibinfo{year}{1969}\natexlab{}.
\newblock \showarticletitle{Quantum detection and estimation theory}.
\newblock \bibinfo{journal}{\emph{Journal of Statistical Physics}} \bibinfo{volume}{1}, \bibinfo{number}{2} (\bibinfo{date}{01 Jun} \bibinfo{year}{1969}), \bibinfo{pages}{231--252}.
\newblock
\showISSN{1572-9613}
\urldef\tempurl%
\url{https://doi.org/10.1007/BF01007479}
\showDOI{\tempurl}


\bibitem[Huang et~al\mbox{.}(2022)]%
        {PCA2}
\bibfield{author}{\bibinfo{person}{Hsin-Yuan Huang}, \bibinfo{person}{Michael Broughton}, \bibinfo{person}{Jordan Cotler}, \bibinfo{person}{Sitan Chen}, \bibinfo{person}{Jerry Li}, \bibinfo{person}{Masoud Mohseni}, \bibinfo{person}{Hartmut Neven}, \bibinfo{person}{Ryan Babbush}, \bibinfo{person}{Richard Kueng}, \bibinfo{person}{John Preskill}, {et~al\mbox{.}}} \bibinfo{year}{2022}\natexlab{}.
\newblock \showarticletitle{Quantum advantage in learning from experiments}.
\newblock \bibinfo{journal}{\emph{Science}} \bibinfo{volume}{376}, \bibinfo{number}{6598} (\bibinfo{date}{jun} \bibinfo{year}{2022}), \bibinfo{pages}{1182--1186}.
\newblock


\bibitem[Huang et~al\mbox{.}(2021)]%
        {PowerdatainQML}
\bibfield{author}{\bibinfo{person}{Hsin-Yuan Huang}, \bibinfo{person}{Michael Broughton}, \bibinfo{person}{Masoud Mohseni}, \bibinfo{person}{Ryan Babbush}, \bibinfo{person}{Sergio Boixo}, \bibinfo{person}{Hartmut Neven}, {and} \bibinfo{person}{Jarrod~R. McClean}.} \bibinfo{year}{2021}\natexlab{}.
\newblock \showarticletitle{Power of data in quantum machine learning}.
\newblock \bibinfo{journal}{\emph{Nature communications}} \bibinfo{volume}{12}, \bibinfo{number}{1} (\bibinfo{date}{feb} \bibinfo{year}{2021}).
\newblock
\urldef\tempurl%
\url{https://doi.org/10.1038/s41467-021-22539-9}
\showDOI{\tempurl}


\bibitem[Huang and Martonosi(2019)]%
        {QSE0.52}
\bibfield{author}{\bibinfo{person}{Yipeng Huang} {and} \bibinfo{person}{Margaret Martonosi}.} \bibinfo{year}{2019}\natexlab{}.
\newblock \showarticletitle{Statistical assertions for validating patterns and finding bugs in quantum programs}. In \bibinfo{booktitle}{\emph{Proceedings of the 46th International Symposium on Computer Architecture}} \emph{(\bibinfo{series}{ISCA '19})}. \bibinfo{publisher}{Association for Computing Machinery}, \bibinfo{address}{New York, NY, USA}, \bibinfo{pages}{541–553}.
\newblock
\urldef\tempurl%
\url{https://doi.org/10.1145/3307650.3322213}
\showDOI{\tempurl}


\bibitem[Humbatova et~al\mbox{.}(2021)]%
        {DeepCrime}
\bibfield{author}{\bibinfo{person}{Nargiz Humbatova}, \bibinfo{person}{Gunel Jahangirova}, {and} \bibinfo{person}{Paolo Tonella}.} \bibinfo{year}{2021}\natexlab{}.
\newblock \showarticletitle{DeepCrime: mutation testing of deep learning systems based on real faults}. In \bibinfo{booktitle}{\emph{Proceedings of the 30th ACM SIGSOFT International Symposium on Software Testing and Analysis}} (Virtual, Denmark) \emph{(\bibinfo{series}{ISSTA 2021})}. \bibinfo{publisher}{Association for Computing Machinery}, \bibinfo{address}{New York, NY, USA}, \bibinfo{pages}{67–78}.
\newblock
\urldef\tempurl%
\url{https://doi.org/10.1145/3460319.3464825}
\showDOI{\tempurl}


\bibitem[Hur et~al\mbox{.}(2022)]%
        {QCNNimg}
\bibfield{author}{\bibinfo{person}{Tak Hur}, \bibinfo{person}{Leeseok Kim}, {and} \bibinfo{person}{Daniel~K. Park}.} \bibinfo{year}{2022}\natexlab{}.
\newblock \showarticletitle{Quantum convolutional neural network for classical data classification}.
\newblock \bibinfo{journal}{\emph{Quantum Machine Intelligence}} \bibinfo{volume}{4}, \bibinfo{number}{1} (\bibinfo{date}{10 Feb} \bibinfo{year}{2022}), \bibinfo{pages}{3}.
\newblock
\showISSN{2524-4914}
\urldef\tempurl%
\url{https://doi.org/10.1007/s42484-021-00061-x}
\showDOI{\tempurl}


\bibitem[Killoran et~al\mbox{.}(2019)]%
        {CVQNN}
\bibfield{author}{\bibinfo{person}{Nathan Killoran}, \bibinfo{person}{Thomas~R. Bromley}, \bibinfo{person}{Juan~Miguel Arrazola}, \bibinfo{person}{Maria Schuld}, \bibinfo{person}{Nicol\'as Quesada}, {and} \bibinfo{person}{Seth Lloyd}.} \bibinfo{year}{2019}\natexlab{}.
\newblock \showarticletitle{Continuous-variable quantum neural networks}.
\newblock \bibinfo{journal}{\emph{Phys. Rev. Res.}}  \bibinfo{volume}{1} (\bibinfo{date}{Oct} \bibinfo{year}{2019}), \bibinfo{pages}{033063}.
\newblock
Issue 3.
\urldef\tempurl%
\url{https://doi.org/10.1103/PhysRevResearch.1.033063}
\showDOI{\tempurl}


\bibitem[Kim et~al\mbox{.}(2019)]%
        {Guidingloss}
\bibfield{author}{\bibinfo{person}{Jinhan Kim}, \bibinfo{person}{Robert Feldt}, {and} \bibinfo{person}{Shin Yoo}.} \bibinfo{year}{2019}\natexlab{}.
\newblock \showarticletitle{Guiding deep learning system testing using surprise adequacy}. In \bibinfo{booktitle}{\emph{2019 IEEE/ACM 41st International Conference on Software Engineering (ICSE)}}. \bibinfo{publisher}{IEEE}, \bibinfo{pages}{1039--1049}.
\newblock


\bibitem[Kurakin et~al\mbox{.}(2018)]%
        {BIM}
\bibfield{author}{\bibinfo{person}{Alexey Kurakin}, \bibinfo{person}{Ian~J Goodfellow}, {and} \bibinfo{person}{Samy Bengio}.} \bibinfo{year}{2018}\natexlab{}.
\newblock \showarticletitle{Adversarial examples in the physical world}.
\newblock In \bibinfo{booktitle}{\emph{Artificial intelligence safety and security}}. \bibinfo{publisher}{Chapman and Hall/CRC}, \bibinfo{pages}{99--112}.
\newblock


\bibitem[LaRose and Coyle(2020)]%
        {larose2020robust}
\bibfield{author}{\bibinfo{person}{Ryan LaRose} {and} \bibinfo{person}{Brian Coyle}.} \bibinfo{year}{2020}\natexlab{}.
\newblock \showarticletitle{Robust data encodings for quantum classifiers}.
\newblock \bibinfo{journal}{\emph{Phys. Rev. A}}  \bibinfo{volume}{102} (\bibinfo{date}{sep} \bibinfo{year}{2020}), \bibinfo{pages}{032420}.
\newblock
Issue 3.
\urldef\tempurl%
\url{https://doi.org/10.1103/PhysRevA.102.032420}
\showDOI{\tempurl}


\bibitem[LeCun et~al\mbox{.}(2010)]%
        {lecun2010mnist}
\bibfield{author}{\bibinfo{person}{Yann LeCun}, \bibinfo{person}{Corinna Cortes}, {and} \bibinfo{person}{CJ Burges}.} \bibinfo{year}{2010}\natexlab{}.
\newblock \showarticletitle{MNIST handwritten digit database}.
\newblock \bibinfo{journal}{\emph{ATT Labs [Online]. Available: http://yann.lecun.com/exdb/mnist}} (\bibinfo{year}{2010}).
\newblock


\bibitem[Li et~al\mbox{.}(2020)]%
        {YMS}
\bibfield{author}{\bibinfo{person}{Gushu Li}, \bibinfo{person}{Li Zhou}, \bibinfo{person}{Nengkun Yu}, \bibinfo{person}{Yufei Ding}, \bibinfo{person}{Mingsheng Ying}, {and} \bibinfo{person}{Yuan Xie}.} \bibinfo{year}{2020}\natexlab{}.
\newblock \showarticletitle{Projection-based runtime assertions for testing and debugging Quantum programs}.
\newblock \bibinfo{journal}{\emph{Proc. ACM Program. Lang.}} \bibinfo{volume}{4}, \bibinfo{number}{OOPSLA}, Article \bibinfo{articleno}{150} (\bibinfo{date}{nov} \bibinfo{year}{2020}), \bibinfo{numpages}{29}~pages.
\newblock
\urldef\tempurl%
\url{https://doi.org/10.1145/3428218}
\showDOI{\tempurl}


\bibitem[Li et~al\mbox{.}(2019)]%
        {QuestionNC_2}
\bibfield{author}{\bibinfo{person}{Zenan Li}, \bibinfo{person}{Xiaoxing Ma}, \bibinfo{person}{Chang Xu}, {and} \bibinfo{person}{Chun Cao}.} \bibinfo{year}{2019}\natexlab{}.
\newblock \showarticletitle{Structural coverage criteria for neural networks could be misleading}. In \bibinfo{booktitle}{\emph{{ICSE} {(NIER)}}} (Montreal, QC, Canada). \bibinfo{publisher}{{IEEE} / {ACM}}, \bibinfo{pages}{89--92}.
\newblock
\urldef\tempurl%
\url{https://doi.org/10.1109/ICSE-NIER.2019.00031}
\showDOI{\tempurl}


\bibitem[Liao et~al\mbox{.}(2021)]%
        {QAML12}
\bibfield{author}{\bibinfo{person}{Haoran Liao}, \bibinfo{person}{Ian Convy}, \bibinfo{person}{William~J. Huggins}, {and} \bibinfo{person}{K.~Birgitta Whaley}.} \bibinfo{year}{2021}\natexlab{}.
\newblock \showarticletitle{Robust in practice: Adversarial attacks on quantum machine learning}.
\newblock \bibinfo{journal}{\emph{Phys. Rev. A}}  \bibinfo{volume}{103} (\bibinfo{date}{Apr} \bibinfo{year}{2021}), \bibinfo{pages}{042427}.
\newblock
Issue 4.
\urldef\tempurl%
\url{https://doi.org/10.1103/PhysRevA.103.042427}
\showDOI{\tempurl}


\bibitem[Liu and Wittek(2020)]%
        {RQD}
\bibfield{author}{\bibinfo{person}{Nana Liu} {and} \bibinfo{person}{Peter Wittek}.} \bibinfo{year}{2020}\natexlab{}.
\newblock \showarticletitle{Vulnerability of quantum classification to adversarial perturbations}.
\newblock \bibinfo{journal}{\emph{Phys. Rev. A}}  \bibinfo{volume}{101} (\bibinfo{date}{jun} \bibinfo{year}{2020}), \bibinfo{pages}{062331}.
\newblock
Issue 6.
\urldef\tempurl%
\url{https://doi.org/10.1103/PhysRevA.101.062331}
\showDOI{\tempurl}


\bibitem[Liu et~al\mbox{.}(2022)]%
        {Deepstate}
\bibfield{author}{\bibinfo{person}{Zixi Liu}, \bibinfo{person}{Yang Feng}, \bibinfo{person}{Yining Yin}, {and} \bibinfo{person}{Zhenyu Chen}.} \bibinfo{year}{2022}\natexlab{}.
\newblock \showarticletitle{DeepState: selecting test suites to enhance the robustness of recurrent neural networks}. In \bibinfo{booktitle}{\emph{Proceedings of the 44th International Conference on Software Engineering}}. \bibinfo{pages}{598--609}.
\newblock


\bibitem[Lloyd et~al\mbox{.}(2014)]%
        {PCA1}
\bibfield{author}{\bibinfo{person}{Seth Lloyd}, \bibinfo{person}{Masoud Mohseni}, {and} \bibinfo{person}{Patrick Rebentrost}.} \bibinfo{year}{2014}\natexlab{}.
\newblock \showarticletitle{Quantum principal component analysis}.
\newblock \bibinfo{journal}{\emph{Nature Physics}} \bibinfo{volume}{10}, \bibinfo{number}{9} (\bibinfo{year}{2014}), \bibinfo{pages}{631--633}.
\newblock


\bibitem[Lu et~al\mbox{.}(2020)]%
        {QAML1}
\bibfield{author}{\bibinfo{person}{Sirui Lu}, \bibinfo{person}{Lu-Ming Duan}, {and} \bibinfo{person}{Dong-Ling Deng}.} \bibinfo{year}{2020}\natexlab{}.
\newblock \showarticletitle{Quantum adversarial machine learning}.
\newblock \bibinfo{journal}{\emph{Physical Review Research}} \bibinfo{volume}{2}, \bibinfo{number}{3} (\bibinfo{date}{jan} \bibinfo{year}{2020}), \bibinfo{pages}{033212}.
\newblock
\urldef\tempurl%
\url{https://doi.org/10.1103/PhysRevResearch.2.033212}
\showDOI{\tempurl}


\bibitem[Ma et~al\mbox{.}(2019a)]%
        {DeepCT}
\bibfield{author}{\bibinfo{person}{Lei Ma}, \bibinfo{person}{Felix Juefei{-}Xu}, \bibinfo{person}{Minhui Xue}, \bibinfo{person}{Bo Li}, \bibinfo{person}{Li Li}, \bibinfo{person}{Yang Liu}, {and} \bibinfo{person}{Jianjun Zhao}.} \bibinfo{year}{2019}\natexlab{a}.
\newblock \showarticletitle{DeepCT: Tomographic Combinatorial Testing for Deep Learning Systems}. In \bibinfo{booktitle}{\emph{26th {IEEE} International Conference on Software Analysis, Evolution and Reengineering, {SANER} 2019, Hangzhou, China, February 24-27, 2019}}, \bibfield{editor}{\bibinfo{person}{Xinyu Wang}, \bibinfo{person}{David Lo}, {and} \bibinfo{person}{Emad Shihab}} (Eds.). \bibinfo{publisher}{{IEEE}}, \bibinfo{pages}{614--618}.
\newblock
\urldef\tempurl%
\url{https://doi.org/10.1109/SANER.2019.8668044}
\showDOI{\tempurl}


\bibitem[Ma et~al\mbox{.}(2018a)]%
        {Deepgauge}
\bibfield{author}{\bibinfo{person}{Lei Ma}, \bibinfo{person}{Felix Juefei-Xu}, \bibinfo{person}{Fuyuan Zhang}, \bibinfo{person}{Jiyuan Sun}, \bibinfo{person}{Minhui Xue}, \bibinfo{person}{Bo Li}, \bibinfo{person}{Chunyang Chen}, \bibinfo{person}{Ting Su}, \bibinfo{person}{Li Li}, \bibinfo{person}{Yang Liu}, {et~al\mbox{.}}} \bibinfo{year}{2018}\natexlab{a}.
\newblock \showarticletitle{Deepgauge: Multi-granularity testing criteria for deep learning systems}. In \bibinfo{booktitle}{\emph{Proceedings of the 33rd ACM/IEEE international conference on automated software engineering}}. \bibinfo{pages}{120--131}.
\newblock


\bibitem[Ma et~al\mbox{.}(2018b)]%
        {DeepMutation}
\bibfield{author}{\bibinfo{person}{Lei Ma}, \bibinfo{person}{Fuyuan Zhang}, \bibinfo{person}{Jiyuan Sun}, \bibinfo{person}{Minhui Xue}, \bibinfo{person}{Bo Li}, \bibinfo{person}{Felix Juefei{-}Xu}, \bibinfo{person}{Chao Xie}, \bibinfo{person}{Li Li}, \bibinfo{person}{Yang Liu}, \bibinfo{person}{Jianjun Zhao}, {and} \bibinfo{person}{Yadong Wang}.} \bibinfo{year}{2018}\natexlab{b}.
\newblock \showarticletitle{DeepMutation: Mutation Testing of Deep Learning Systems}. In \bibinfo{booktitle}{\emph{2018 IEEE 29th international symposium on software reliability engineering (ISSRE)}} (Memphis, TN, USA). \bibinfo{publisher}{{IEEE} Computer Society}, \bibinfo{pages}{100--111}.
\newblock
\urldef\tempurl%
\url{https://doi.org/10.1109/ISSRE.2018.00021}
\showDOI{\tempurl}


\bibitem[Ma et~al\mbox{.}(2019b)]%
        {ma2019paddlepaddle}
\bibfield{author}{\bibinfo{person}{Yanjun Ma}, \bibinfo{person}{Dianhai Yu}, \bibinfo{person}{Tian Wu}, {and} \bibinfo{person}{Haifeng Wang}.} \bibinfo{year}{2019}\natexlab{b}.
\newblock \showarticletitle{PaddlePaddle: An open-source deep learning platform from industrial practice}.
\newblock \bibinfo{journal}{\emph{Frontiers of Data and Domputing}} \bibinfo{volume}{1}, \bibinfo{number}{1} (\bibinfo{year}{2019}), \bibinfo{pages}{105--115}.
\newblock


\bibitem[Madry et~al\mbox{.}(2017)]%
        {PGD}
\bibfield{author}{\bibinfo{person}{Aleksander Madry}, \bibinfo{person}{Aleksandar Makelov}, \bibinfo{person}{Ludwig Schmidt}, \bibinfo{person}{Dimitris Tsipras}, {and} \bibinfo{person}{Adrian Vladu}.} \bibinfo{year}{2017}\natexlab{}.
\newblock \showarticletitle{Towards deep learning models resistant to adversarial attacks}.
\newblock  (\bibinfo{year}{2017}).
\newblock
\showeprint{arXiv preprint arXiv:1706.06083}


\bibitem[Mendiluze et~al\mbox{.}(2021)]%
        {QSE2}
\bibfield{author}{\bibinfo{person}{E{\~n}aut Mendiluze}, \bibinfo{person}{Shaukat Ali}, \bibinfo{person}{Paolo Arcaini}, {and} \bibinfo{person}{Tao Yue}.} \bibinfo{year}{2021}\natexlab{}.
\newblock \showarticletitle{Muskit: A mutation analysis tool for quantum software testing}. In \bibinfo{booktitle}{\emph{2021 36th IEEE/ACM International Conference on Automated Software Engineering (ASE)}}. \bibinfo{publisher}{IEEE}, \bibinfo{pages}{1266--1270}.
\newblock


\bibitem[Meyer and Wallach(2002)]%
        {entanglement}
\bibfield{author}{\bibinfo{person}{David~A. Meyer} {and} \bibinfo{person}{Nolan~R. Wallach}.} \bibinfo{year}{2002}\natexlab{}.
\newblock \showarticletitle{Global entanglement in multiparticle systems}.
\newblock \bibinfo{journal}{\emph{J. Math. Phys.}} \bibinfo{volume}{43}, \bibinfo{number}{9} (\bibinfo{date}{aug} \bibinfo{year}{2002}), \bibinfo{pages}{4273--4278}.
\newblock
\showISSN{0022-2488}
\urldef\tempurl%
\url{https://doi.org/10.1063/1.1497700}
\showDOI{\tempurl}


\bibitem[Miranskyy and Zhang(2019)]%
        {QSE0.51}
\bibfield{author}{\bibinfo{person}{Andriy Miranskyy} {and} \bibinfo{person}{Lei Zhang}.} \bibinfo{year}{2019}\natexlab{}.
\newblock \showarticletitle{On testing quantum programs}. In \bibinfo{booktitle}{\emph{Proceedings of the 41st International Conference on Software Engineering: New Ideas and Emerging Results}} (Montreal, Quebec, Canada) \emph{(\bibinfo{series}{ICSE-NIER '19})}. \bibinfo{publisher}{IEEE Press}, \bibinfo{pages}{57–60}.
\newblock
\urldef\tempurl%
\url{https://doi.org/10.1109/ICSE-NIER.2019.00023}
\showDOI{\tempurl}


\bibitem[Mitarai et~al\mbox{.}(2018)]%
        {mitarai2018quantum}
\bibfield{author}{\bibinfo{person}{K. Mitarai}, \bibinfo{person}{M. Negoro}, \bibinfo{person}{M. Kitagawa}, {and} \bibinfo{person}{K. Fujii}.} \bibinfo{year}{2018}\natexlab{}.
\newblock \showarticletitle{Quantum circuit learning}.
\newblock \bibinfo{journal}{\emph{Phys. Rev. A}}  \bibinfo{volume}{98} (\bibinfo{date}{sep} \bibinfo{year}{2018}), \bibinfo{pages}{032309}.
\newblock
Issue 3.
\urldef\tempurl%
\url{https://doi.org/10.1103/PhysRevA.98.032309}
\showDOI{\tempurl}


\bibitem[Moosavi{-}Dezfooli et~al\mbox{.}(2016)]%
        {DeepFool}
\bibfield{author}{\bibinfo{person}{Seyed{-}Mohsen Moosavi{-}Dezfooli}, \bibinfo{person}{Alhussein Fawzi}, {and} \bibinfo{person}{Pascal Frossard}.} \bibinfo{year}{2016}\natexlab{}.
\newblock \showarticletitle{DeepFool: {A} Simple and Accurate Method to Fool Deep Neural Networks}. In \bibinfo{booktitle}{\emph{{CVPR}}}. \bibinfo{publisher}{{IEEE} Computer Society}, \bibinfo{pages}{2574--2582}.
\newblock


\bibitem[Muqeet et~al\mbox{.}(2023)]%
        {QSE3}
\bibfield{author}{\bibinfo{person}{Asmar Muqeet}, \bibinfo{person}{Tao Yue}, \bibinfo{person}{Shaukat Ali}, {and} \bibinfo{person}{Paolo Arcaini}.} \bibinfo{year}{2023}\natexlab{}.
\newblock \showarticletitle{Noise-Aware Quantum Software Testing}.
\newblock  (\bibinfo{year}{2023}).
\newblock
\urldef\tempurl%
\url{https://doi.org/10.48550/arXiv.2306.16992}
\showDOI{\tempurl}
\showeprint{arXiv preprint arXiv:2306.16992}


\bibitem[Nguyen et~al\mbox{.}(2015)]%
        {Advattention1}
\bibfield{author}{\bibinfo{person}{Anh~Mai Nguyen}, \bibinfo{person}{Jason Yosinski}, {and} \bibinfo{person}{Jeff Clune}.} \bibinfo{year}{2015}\natexlab{}.
\newblock \showarticletitle{Deep neural networks are easily fooled: High confidence predictions for unrecognizable images}. In \bibinfo{booktitle}{\emph{{CVPR}}}. \bibinfo{publisher}{{IEEE} Computer Society}, \bibinfo{pages}{427--436}.
\newblock


\bibitem[Nielsen and Chuang(2010)]%
        {QCQI}
\bibfield{author}{\bibinfo{person}{Michael~A Nielsen} {and} \bibinfo{person}{Isaac~L Chuang}.} \bibinfo{year}{2010}\natexlab{}.
\newblock \bibinfo{booktitle}{\emph{Quantum computation and quantum information}}.
\newblock \bibinfo{publisher}{Cambridge university press}.
\newblock
\urldef\tempurl%
\url{https://doi.org/10.1017/CBO9780511976667}
\showDOI{\tempurl}


\bibitem[Oh et~al\mbox{.}(2019)]%
        {Fidelity}
\bibfield{author}{\bibinfo{person}{Changhun Oh}, \bibinfo{person}{Changhyoup Lee}, \bibinfo{person}{Leonardo Banchi}, \bibinfo{person}{Su-Yong Lee}, \bibinfo{person}{Carsten Rockstuhl}, {and} \bibinfo{person}{Hyunseok Jeong}.} \bibinfo{year}{2019}\natexlab{}.
\newblock \showarticletitle{Optimal measurements for quantum fidelity between Gaussian states and its relevance to quantum metrology}.
\newblock \bibinfo{journal}{\emph{Phys. Rev. A}}  \bibinfo{volume}{100} (\bibinfo{date}{Jul} \bibinfo{year}{2019}), \bibinfo{pages}{012323}.
\newblock
Issue 1.
\urldef\tempurl%
\url{https://doi.org/10.1103/PhysRevA.100.012323}
\showDOI{\tempurl}


\bibitem[Papernot et~al\mbox{.}(2016a)]%
        {JSMA}
\bibfield{author}{\bibinfo{person}{Nicolas Papernot}, \bibinfo{person}{Patrick McDaniel}, \bibinfo{person}{Somesh Jha}, \bibinfo{person}{Matt Fredrikson}, \bibinfo{person}{Z~Berkay Celik}, {and} \bibinfo{person}{Ananthram Swami}.} \bibinfo{year}{2016}\natexlab{a}.
\newblock \showarticletitle{The limitations of deep learning in adversarial settings}. In \bibinfo{booktitle}{\emph{2016 IEEE European symposium on security and privacy (EuroS\&P)}}. \bibinfo{publisher}{IEEE}, \bibinfo{pages}{372--387}.
\newblock


\bibitem[Papernot et~al\mbox{.}(2016b)]%
        {attackTEXT}
\bibfield{author}{\bibinfo{person}{Nicolas Papernot}, \bibinfo{person}{Patrick~D. McDaniel}, \bibinfo{person}{Ananthram Swami}, {and} \bibinfo{person}{Richard~E. Harang}.} \bibinfo{year}{2016}\natexlab{b}.
\newblock \showarticletitle{Crafting adversarial input sequences for recurrent neural networks}. In \bibinfo{booktitle}{\emph{MILCOM 2016-2016 IEEE Military Communications Conference}}. \bibinfo{publisher}{{IEEE}}, \bibinfo{pages}{49--54}.
\newblock


\bibitem[Pei et~al\mbox{.}(2017)]%
        {Deepxplore}
\bibfield{author}{\bibinfo{person}{Kexin Pei}, \bibinfo{person}{Yinzhi Cao}, \bibinfo{person}{Junfeng Yang}, {and} \bibinfo{person}{Suman Jana}.} \bibinfo{year}{2017}\natexlab{}.
\newblock \showarticletitle{Deepxplore: Automated whitebox testing of deep learning systems}. In \bibinfo{booktitle}{\emph{proceedings of the 26th Symposium on Operating Systems Principles}}. \bibinfo{pages}{1--18}.
\newblock


\bibitem[Pesah et~al\mbox{.}(2021)]%
        {pesah2021absence}
\bibfield{author}{\bibinfo{person}{Arthur Pesah}, \bibinfo{person}{M. Cerezo}, \bibinfo{person}{Samson Wang}, \bibinfo{person}{Tyler Volkoff}, \bibinfo{person}{Andrew~T. Sornborger}, {and} \bibinfo{person}{Patrick~J. Coles}.} \bibinfo{year}{2021}\natexlab{}.
\newblock \showarticletitle{Absence of Barren Plateaus in Quantum Convolutional Neural Networks}.
\newblock \bibinfo{journal}{\emph{Phys. Rev. X}}  \bibinfo{volume}{11} (\bibinfo{date}{oct} \bibinfo{year}{2021}), \bibinfo{pages}{041011}.
\newblock
Issue 4.
\urldef\tempurl%
\url{https://doi.org/10.1103/PhysRevX.11.041011}
\showDOI{\tempurl}


\bibitem[Peters et~al\mbox{.}(2021)]%
        {QKM}
\bibfield{author}{\bibinfo{person}{Evan Peters}, \bibinfo{person}{Jo{\~a}o Caldeira}, \bibinfo{person}{Alan Ho}, \bibinfo{person}{Stefan Leichenauer}, \bibinfo{person}{Masoud Mohseni}, \bibinfo{person}{Hartmut Neven}, \bibinfo{person}{Panagiotis Spentzouris}, \bibinfo{person}{Doug Strain}, {and} \bibinfo{person}{Gabriel~N Perdue}.} \bibinfo{year}{2021}\natexlab{}.
\newblock \showarticletitle{Machine learning of high dimensional data on a noisy quantum processor}.
\newblock \bibinfo{journal}{\emph{npj Quantum Information}} \bibinfo{volume}{7}, \bibinfo{number}{1} (\bibinfo{date}{nov} \bibinfo{year}{2021}), \bibinfo{pages}{161}.
\newblock
\urldef\tempurl%
\url{https://doi.org/10.1038/s41534-021-00498-9}
\showDOI{\tempurl}


\bibitem[Ren et~al\mbox{.}(2022)]%
        {QAML2}
\bibfield{author}{\bibinfo{person}{Wenhui Ren}, \bibinfo{person}{Weikang Li}, \bibinfo{person}{Shibo Xu}, \bibinfo{person}{Ke Wang}, \bibinfo{person}{Wenjie Jiang}, \bibinfo{person}{Feitong Jin}, \bibinfo{person}{Xuhao Zhu}, \bibinfo{person}{Jiachen Chen}, \bibinfo{person}{Zixuan Song}, \bibinfo{person}{Pengfei Zhang}, {et~al\mbox{.}}} \bibinfo{year}{2022}\natexlab{}.
\newblock \showarticletitle{Experimental quantum adversarial learning with programmable superconducting qubits}.
\newblock \bibinfo{journal}{\emph{Nature Computational Science}} \bibinfo{volume}{2}, \bibinfo{number}{11} (\bibinfo{date}{apr} \bibinfo{year}{2022}), \bibinfo{pages}{711--717}.
\newblock
\urldef\tempurl%
\url{https://doi.org/10.1038/s43588-022-00351-9}
\showDOI{\tempurl}


\bibitem[Schuld(2021)]%
        {schuld2021supervised}
\bibfield{author}{\bibinfo{person}{Maria Schuld}.} \bibinfo{year}{2021}\natexlab{}.
\newblock \bibinfo{title}{Supervised quantum machine learning models are kernel methods}.
\newblock
\newblock
\urldef\tempurl%
\url{https://doi.org/10.48550/arXiv.2101.11020}
\showDOI{\tempurl}
\showeprint[arxiv]{2101.11020}~[quant-ph]


\bibitem[Schuld et~al\mbox{.}(2020)]%
        {schuld2020circuit}
\bibfield{author}{\bibinfo{person}{Maria Schuld}, \bibinfo{person}{Alex Bocharov}, \bibinfo{person}{Krysta~M. Svore}, {and} \bibinfo{person}{Nathan Wiebe}.} \bibinfo{year}{2020}\natexlab{}.
\newblock \showarticletitle{Circuit-centric quantum classifiers}.
\newblock \bibinfo{journal}{\emph{Phys. Rev. A}}  \bibinfo{volume}{101} (\bibinfo{date}{mar} \bibinfo{year}{2020}), \bibinfo{pages}{032308}.
\newblock
Issue 3.
\urldef\tempurl%
\url{https://doi.org/10.1103/PhysRevA.101.032308}
\showDOI{\tempurl}


\bibitem[Schuld and Petruccione(2018)]%
        {schuld2018supervised}
\bibfield{author}{\bibinfo{person}{Maria Schuld} {and} \bibinfo{person}{Francesco Petruccione}.} \bibinfo{year}{2018}\natexlab{}.
\newblock \bibinfo{booktitle}{\emph{Information Encoding}}.
\newblock \bibinfo{publisher}{Springer International Publishing}, \bibinfo{address}{Cham}, \bibinfo{pages}{139--171}.
\newblock
\showISBNx{978-3-319-96424-9}
\urldef\tempurl%
\url{https://doi.org/10.1007/978-3-319-96424-9_5}
\showDOI{\tempurl}


\bibitem[Sharif et~al\mbox{.}(2016)]%
        {attackIMG}
\bibfield{author}{\bibinfo{person}{Mahmood Sharif}, \bibinfo{person}{Sruti Bhagavatula}, \bibinfo{person}{Lujo Bauer}, {and} \bibinfo{person}{Michael~K. Reiter}.} \bibinfo{year}{2016}\natexlab{}.
\newblock \showarticletitle{Accessorize to a Crime: Real and Stealthy Attacks on State-of-the-Art Face Recognition}. In \bibinfo{booktitle}{\emph{Proceedings of the 2016 acm sigsac conference on computer and communications security}}. \bibinfo{publisher}{{ACM}}, \bibinfo{pages}{1528--1540}.
\newblock


\bibitem[Shi et~al\mbox{.}(2022)]%
        {PHL}
\bibfield{author}{\bibinfo{person}{Jinjing Shi}, \bibinfo{person}{Wenxuan Wang}, \bibinfo{person}{Xiaoping Lou}, \bibinfo{person}{Shichao Zhang}, {and} \bibinfo{person}{Xuelong Li}.} \bibinfo{year}{2022}\natexlab{}.
\newblock \showarticletitle{Parameterized Hamiltonian learning with quantum circuit}.
\newblock \bibinfo{journal}{\emph{IEEE Transactions on Pattern Analysis and Machine Intelligence}} \bibinfo{volume}{45}, \bibinfo{number}{5} (\bibinfo{date}{Aug} \bibinfo{year}{2022}), \bibinfo{pages}{6086--6095}.
\newblock


\bibitem[Sim et~al\mbox{.}(2019)]%
        {ExpEnt}
\bibfield{author}{\bibinfo{person}{Sukin Sim}, \bibinfo{person}{Peter~D Johnson}, {and} \bibinfo{person}{Al{\'a}n Aspuru-Guzik}.} \bibinfo{year}{2019}\natexlab{}.
\newblock \showarticletitle{Expressibility and entangling capability of parameterized quantum circuits for hybrid quantum-classical algorithms}.
\newblock \bibinfo{journal}{\emph{Advanced Quantum Technologies}} \bibinfo{volume}{2}, \bibinfo{number}{12} (\bibinfo{year}{2019}), \bibinfo{pages}{1900070}.
\newblock
\urldef\tempurl%
\url{https://doi.org/10.1002/qute.201900070}
\showDOI{\tempurl}


\bibitem[Steinbrecher et~al\mbox{.}(2019)]%
        {QONN}
\bibfield{author}{\bibinfo{person}{Gregory~R. Steinbrecher}, \bibinfo{person}{Jonathan~P. Olson}, \bibinfo{person}{Dirk Englund}, {and} \bibinfo{person}{Jacques Carolan}.} \bibinfo{year}{2019}\natexlab{}.
\newblock \showarticletitle{Quantum optical neural networks}.
\newblock \bibinfo{journal}{\emph{npj Quantum Information}} \bibinfo{volume}{5}, \bibinfo{number}{1} (\bibinfo{date}{17 Jul} \bibinfo{year}{2019}), \bibinfo{pages}{60}.
\newblock
\showISSN{2056-6387}
\urldef\tempurl%
\url{https://doi.org/10.1038/s41534-019-0174-7}
\showDOI{\tempurl}


\bibitem[Szegedy et~al\mbox{.}(2013)]%
        {14ADV}
\bibfield{author}{\bibinfo{person}{Christian Szegedy}, \bibinfo{person}{Wojciech Zaremba}, \bibinfo{person}{Ilya Sutskever}, \bibinfo{person}{Joan Bruna}, \bibinfo{person}{Dumitru Erhan}, \bibinfo{person}{Ian Goodfellow}, {and} \bibinfo{person}{Rob Fergus}.} \bibinfo{year}{2013}\natexlab{}.
\newblock \showarticletitle{Intriguing properties of neural networks}.
\newblock  (\bibinfo{year}{2013}).
\newblock
\urldef\tempurl%
\url{https://doi.org/10.48550/arXiv.1312.6199}
\showDOI{\tempurl}
\showeprint{arXiv preprint arXiv:1312.6199}


\bibitem[Szegedy et~al\mbox{.}(2014)]%
        {Advattention2}
\bibfield{author}{\bibinfo{person}{Christian Szegedy}, \bibinfo{person}{Wojciech Zaremba}, \bibinfo{person}{Ilya Sutskever}, \bibinfo{person}{Joan Bruna}, \bibinfo{person}{Dumitru Erhan}, \bibinfo{person}{Ian~J. Goodfellow}, {and} \bibinfo{person}{Rob Fergus}.} \bibinfo{year}{2014}\natexlab{}.
\newblock \showarticletitle{Intriguing properties of neural networks}. In \bibinfo{booktitle}{\emph{{ICLR} (Poster)}}.
\newblock


\bibitem[Tian et~al\mbox{.}(2018)]%
        {Deeptest}
\bibfield{author}{\bibinfo{person}{Yuchi Tian}, \bibinfo{person}{Kexin Pei}, \bibinfo{person}{Suman Jana}, {and} \bibinfo{person}{Baishakhi Ray}.} \bibinfo{year}{2018}\natexlab{}.
\newblock \showarticletitle{Deeptest: Automated testing of deep-neural-network-driven autonomous cars}. In \bibinfo{booktitle}{\emph{Proceedings of the 40th international conference on software engineering}}. \bibinfo{pages}{303--314}.
\newblock


\bibitem[Wang et~al\mbox{.}(2021a)]%
        {Robot}
\bibfield{author}{\bibinfo{person}{Jingyi Wang}, \bibinfo{person}{Jialuo Chen}, \bibinfo{person}{Youcheng Sun}, \bibinfo{person}{Xingjun Ma}, \bibinfo{person}{Dongxia Wang}, \bibinfo{person}{Jun Sun}, {and} \bibinfo{person}{Peng Cheng}.} \bibinfo{year}{2021}\natexlab{a}.
\newblock \showarticletitle{RobOT: Robustness-oriented testing for deep learning systems}. In \bibinfo{booktitle}{\emph{2021 IEEE/ACM 43rd International Conference on Software Engineering (ICSE)}}. \bibinfo{publisher}{IEEE}, \bibinfo{pages}{300--311}.
\newblock


\bibitem[Wang et~al\mbox{.}(2019)]%
        {Mutation2}
\bibfield{author}{\bibinfo{person}{Jingyi Wang}, \bibinfo{person}{Guoliang Dong}, \bibinfo{person}{Jun Sun}, \bibinfo{person}{Xinyu Wang}, {and} \bibinfo{person}{Peixin Zhang}.} \bibinfo{year}{2019}\natexlab{}.
\newblock \showarticletitle{Adversarial sample detection for deep neural network through model mutation testing}. In \bibinfo{booktitle}{\emph{Proceedings of the 41st International Conference on Software Engineering}} (Montreal, Quebec, Canada) \emph{(\bibinfo{series}{ICSE '19})}. \bibinfo{publisher}{IEEE Press}, \bibinfo{pages}{1245–1256}.
\newblock
\urldef\tempurl%
\url{https://doi.org/10.1109/ICSE.2019.00126}
\showDOI{\tempurl}


\bibitem[Wang et~al\mbox{.}(2021b)]%
        {QSE1}
\bibfield{author}{\bibinfo{person}{Jiyuan Wang}, \bibinfo{person}{Fucheng Ma}, {and} \bibinfo{person}{Yu Jiang}.} \bibinfo{year}{2021}\natexlab{b}.
\newblock \showarticletitle{Poster: Fuzz testing of quantum program}. In \bibinfo{booktitle}{\emph{2021 14th IEEE Conference on Software Testing, Verification and Validation (ICST)}}. \bibinfo{publisher}{IEEE}, \bibinfo{pages}{466--469}.
\newblock


\bibitem[Wang et~al\mbox{.}(2022b)]%
        {QSEdiff}
\bibfield{author}{\bibinfo{person}{Jiyuan Wang}, \bibinfo{person}{Qian Zhang}, \bibinfo{person}{Guoqing~Harry Xu}, {and} \bibinfo{person}{Miryung Kim}.} \bibinfo{year}{2022}\natexlab{b}.
\newblock \showarticletitle{QDiff: differential testing of quantum software stacks}. In \bibinfo{booktitle}{\emph{Proceedings of the 36th IEEE/ACM International Conference on Automated Software Engineering}} (Melbourne, Australia) \emph{(\bibinfo{series}{ASE '21})}. \bibinfo{publisher}{IEEE Press}, \bibinfo{pages}{692–704}.
\newblock
\urldef\tempurl%
\url{https://doi.org/10.1109/ASE51524.2021.9678792}
\showDOI{\tempurl}


\bibitem[Wang et~al\mbox{.}(2022a)]%
        {QSEcov1}
\bibfield{author}{\bibinfo{person}{Xinyi Wang}, \bibinfo{person}{Paolo Arcaini}, \bibinfo{person}{Tao Yue}, {and} \bibinfo{person}{Shaukat Ali}.} \bibinfo{year}{2022}\natexlab{a}.
\newblock \showarticletitle{Quito: a coverage-guided test generator for quantum programs}. In \bibinfo{booktitle}{\emph{Proceedings of the 36th IEEE/ACM International Conference on Automated Software Engineering}} (Melbourne, Australia) \emph{(\bibinfo{series}{ASE '21})}. \bibinfo{publisher}{IEEE Press}, \bibinfo{pages}{1237–1241}.
\newblock
\urldef\tempurl%
\url{https://doi.org/10.1109/ASE51524.2021.9678798}
\showDOI{\tempurl}


\bibitem[Xiao et~al\mbox{.}(2017)]%
        {DBLP:journals/corr/abs-1708-07747}
\bibfield{author}{\bibinfo{person}{Han Xiao}, \bibinfo{person}{Kashif Rasul}, {and} \bibinfo{person}{Roland Vollgraf}.} \bibinfo{year}{2017}\natexlab{}.
\newblock \showarticletitle{Fashion-MNIST: a Novel Image Dataset for Benchmarking Machine Learning Algorithms}.
\newblock \bibinfo{journal}{\emph{CoRR}}  \bibinfo{volume}{abs/1708.07747} (\bibinfo{year}{2017}).
\newblock
\showeprint[arxiv]{1708.07747}
\urldef\tempurl%
\url{http://arxiv.org/abs/1708.07747}
\showURL{%
\tempurl}


\end{thebibliography}

\end{document}